\def\PsfigVersion{1.10}
\def\setDriver{\DvipsDriver} 
\ifx\undefined\psfig\else \fi
\let\LaTeXAtSign=\@
\let\@=\relax
\edef\psfigRestoreAt{\catcode`\@=\number\catcode`@\relax}
\catcode`\@=11\relax
\newwrite\@unused
\def\ps@typeout#1{{\let\protect\string\immediate\write\@unused{#1}}}

\def\DvipsDriver{
	\ps@typeout{psfig/tex \PsfigVersion -dvips}
\def\PsfigSpecials{\DvipsSpecials} 	\def\ps@dir{/}
\def\ps@predir{} }
\def\OzTeXDriver{
	\ps@typeout{psfig/tex \PsfigVersion -oztex}
	\def\PsfigSpecials{\OzTeXSpecials}
	\def\ps@dir{:}
	\def\ps@predir{:}
	\catcode`\^^J=5
}

\def\figurepath{./:}

\def\DoPaths#1{\expandafter\EachPath#1\stoplist}
\def\leer{}
\def\EachPath#1:#2\stoplist{
  \ExistsFile{#1}{\SearchedFile}
  \ifx#2\leer
  \else
    \expandafter\EachPath#2\stoplist
  \fi}
\def\ps@dir{/}
\def\ExistsFile#1#2{%
   \openin1=\ps@predir#1\ps@dir#2
   \ifeof1
       \closein1
   \else
       \closein1
        \ifx\ps@founddir\leer
           \edef\ps@founddir{#1}
        \fi
   \fi}
\def\get@dir#1{%
  \def\ps@founddir{}
  \def\SearchedFile{#1}
  \DoPaths\figurepath
}

\def\@nnil{\@nil}
\def\@empty{}
\def\@psdonoop#1\@@#2#3{}
\def\@psdo#1:=#2\do#3{\edef\@psdotmp{#2}\ifx\@psdotmp\@empty \else
    \expandafter\@psdoloop#2,\@nil,\@nil\@@#1{#3}\fi}
\def\@psdoloop#1,#2,#3\@@#4#5{\def#4{#1}\ifx #4\@nnil \else
       #5\def#4{#2}\ifx #4\@nnil \else#5\@ipsdoloop #3\@@#4{#5}\fi\fi}
\def\@ipsdoloop#1,#2\@@#3#4{\def#3{#1}\ifx #3\@nnil 
       \let\@nextwhile=\@psdonoop \else
      #4\relax\let\@nextwhile=\@ipsdoloop\fi\@nextwhile#2\@@#3{#4}}
\def\@tpsdo#1:=#2\do#3{\xdef\@psdotmp{#2}\ifx\@psdotmp\@empty \else
    \@tpsdoloop#2\@nil\@nil\@@#1{#3}\fi}
\def\@tpsdoloop#1#2\@@#3#4{\def#3{#1}\ifx #3\@nnil 
       \let\@nextwhile=\@psdonoop \else
      #4\relax\let\@nextwhile=\@tpsdoloop\fi\@nextwhile#2\@@#3{#4}}
\ifx\undefined\fbox
\newdimen\fboxrule
\newdimen\fboxsep
\newdimen\ps@tempdima
\newbox\ps@tempboxa
\long\def\fbox#1{\leavevmode\setbox\ps@tempboxa\hbox{#1}\ps@tempdima\fboxrule
    \advance\ps@tempdima \fboxsep \advance\ps@tempdima \dp\ps@tempboxa
   \hbox{\lower \ps@tempdima\hbox
  {\vbox{\hrule height \fboxrule
          \hbox{\vrule width \fboxrule \hskip\fboxsep
          \vbox{\vskip\fboxsep \box\ps@tempboxa\vskip\fboxsep}\hskip 
                 \fboxsep\vrule width \fboxrule}
                 \hrule height \fboxrule}}}}
\fi
\newread\ps@stream
\newif\ifnot@eof       
\newif\if@noisy        
\newif\if@atend        
\newif\if@psfile       
{\catcode`\%=12\global\gdef\epsf@start{
\def\epsf@PS{PS}
\def\epsf@getbb#1{%
\openin\ps@stream=\ps@predir#1
\ifeof\ps@stream\ps@typeout{Error, File #1 not found}\else
   {\not@eoftrue \chardef\other=12
    \def\do##1{\catcode`##1=\other}\dospecials \catcode`\ =10
    \loop
       \if@psfile
	  \read\ps@stream to \epsf@fileline
       \else{
	  \obeyspaces
          \read\ps@stream to \epsf@tmp\global\let\epsf@fileline\epsf@tmp}
       \fi
       \ifeof\ps@stream\not@eoffalse\else
       \if@psfile\else
       \expandafter\epsf@test\epsf@fileline:. \\%
       \fi
          \expandafter\epsf@aux\epsf@fileline:. \\%
       \fi
   \ifnot@eof\repeat
   }\closein\ps@stream\fi}%
\long\def\epsf@test#1#2#3:#4\\{\def\epsf@testit{#1#2}
			\ifx\epsf@testit\epsf@start\else
\ps@typeout{Warning! File does not start with `\epsf@start'.  It may not be a PostScript file.}
			\fi
			\@psfiletrue} 
{\catcode`\%=12\global\let\epsf@percent=
\long\def\epsf@aux#1#2:#3\\{\ifx#1\epsf@percent
   \def\epsf@testit{#2}\ifx\epsf@testit\epsf@bblit
	\@atendfalse
        \epsf@atend #3 . \\%
	\if@atend	
	   \if@verbose{
		\ps@typeout{psfig: found `(atend)'; continuing search}
	   }\fi
        \else
        \epsf@grab #3 . . . \\%
        \not@eoffalse
        \global\no@bbfalse
        \fi
   \fi\fi}%
\def\epsf@grab #1 #2 #3 #4 #5\\{%
   \global\def\epsf@llx{#1}\ifx\epsf@llx\empty
      \epsf@grab #2 #3 #4 #5 .\\\else
   \global\def\epsf@lly{#2}%
   \global\def\epsf@urx{#3}\global\def\epsf@ury{#4}\fi}%
\def\epsf@atendlit{(atend)} 
\def\epsf@atend #1 #2 #3\\{%
   \def\epsf@tmp{#1}\ifx\epsf@tmp\empty
      \epsf@atend #2 #3 .\\\else
   \ifx\epsf@tmp\epsf@atendlit\@atendtrue\fi\fi}

\chardef\psletter = 11 
\chardef\other = 12

\newif \ifdebug 
\newif\ifc@mpute 
\c@mputetrue 

\let\then = \relax
\def\r@dian{pt }
\let\r@dians = \r@dian
\let\dimensionless@nit = \r@dian
\let\dimensionless@nits = \dimensionless@nit
\def\internal@nit{sp }
\let\internal@nits = \internal@nit
\newif\ifstillc@nverging
\def \Mess@ge #1{\ifdebug \then \message {#1} \fi}

{ 
	\catcode `\@ = \psletter
	\gdef \nodimen {\expandafter \n@dimen \the \dimen}
	\gdef \term #1 #2 #3%
	       {\edef \t@ {\the #1}
		\edef \t@@ {\expandafter \n@dimen \the #2\r@dian}%
		\t@rm {\t@} {\t@@} {#3}%
	       }
	\gdef \t@rm #1 #2 #3%
	       {{%
		\count 0 = 0
		\dimen 0 = 1 \dimensionless@nit
		\dimen 2 = #2\relax
		\Mess@ge {Calculating term #1 of \nodimen 2}%
		\loop
		\ifnum	\count 0 < #1
		\then	\advance \count 0 by 1
			\Mess@ge {Iteration \the \count 0 \space}%
			\Multiply \dimen 0 by {\dimen 2}%
			\Mess@ge {After multiplication, term = \nodimen 0}%
			\Divide \dimen 0 by {\count 0}%
			\Mess@ge {After division, term = \nodimen 0}%
		\repeat
		\Mess@ge {Final value for term #1 of 
				\nodimen 2 \space is \nodimen 0}%
		\xdef \Term {#3 = \nodimen 0 \r@dians}%
		\aftergroup \Term
	       }}
	\catcode `\p = \other
	\catcode `\t = \other
	\gdef \n@dimen #1pt{#1} 
}

\def \Divide #1by #2{\divide #1 by #2} 

\def \Multiply #1by #2
       {{
	\count 0 = #1\relax
	\count 2 = #2\relax
	\count 4 = 65536
	\Mess@ge {Before scaling, count 0 = \the \count 0 \space and
			count 2 = \the \count 2}%
	\ifnum	\count 0 > 32767 
	\then	\divide \count 0 by 4
		\divide \count 4 by 4
	\else	\ifnum	\count 0 < -32767
		\then	\divide \count 0 by 4
			\divide \count 4 by 4
		\else
		\fi
	\fi
	\ifnum	\count 2 > 32767 
	\then	\divide \count 2 by 4
		\divide \count 4 by 4
	\else	\ifnum	\count 2 < -32767
		\then	\divide \count 2 by 4
			\divide \count 4 by 4
		\else
		\fi
	\fi
	\multiply \count 0 by \count 2
	\divide \count 0 by \count 4
	\xdef \product {#1 = \the \count 0 \internal@nits}%
	\aftergroup \product
       }}

\def\r@duce{\ifdim\dimen0 > 90\r@dian \then   
		\multiply\dimen0 by -1
		\advance\dimen0 by 180\r@dian
		\r@duce
	    \else \ifdim\dimen0 < -90\r@dian \then  
		\advance\dimen0 by 360\r@dian
		\r@duce
		\fi
	    \fi}

\def\Sine#1%
       {{%
	\dimen 0 = #1 \r@dian
	\r@duce
	\ifdim\dimen0 = -90\r@dian \then
	   \dimen4 = -1\r@dian
	   \c@mputefalse
	\fi
	\ifdim\dimen0 = 90\r@dian \then
	   \dimen4 = 1\r@dian
	   \c@mputefalse
	\fi
	\ifdim\dimen0 = 0\r@dian \then
	   \dimen4 = 0\r@dian
	   \c@mputefalse
	\fi
	\ifc@mpute \then
		\divide\dimen0 by 180
		\dimen0=3.141592654\dimen0
		\dimen 2 = 3.1415926535897963\r@dian 
		\divide\dimen 2 by 2 
		\Mess@ge {Sin: calculating Sin of \nodimen 0}%
		\count 0 = 1 
		\dimen 2 = 1 \r@dian 
		\dimen 4 = 0 \r@dian 
		\loop
			\ifnum	\dimen 2 = 0 
			\then	\stillc@nvergingfalse 
			\else	\stillc@nvergingtrue
			\fi
			\ifstillc@nverging 
			\then	\term {\count 0} {\dimen 0} {\dimen 2}%
				\advance \count 0 by 2
				\count 2 = \count 0
				\divide \count 2 by 2
				\ifodd	\count 2 
				\then	\advance \dimen 4 by \dimen 2
				\else	\advance \dimen 4 by -\dimen 2
				\fi
		\repeat
	\fi		
			\xdef \sine {\nodimen 4}%
       }}

\def\Cosine#1{\ifx\sine\UnDefined\edef\Savesine{\relax}\else
		             \edef\Savesine{\sine}\fi
	{\dimen0=#1\r@dian\advance\dimen0 by 90\r@dian
	 \Sine{\nodimen 0}
	 \xdef\cosine{\sine}
	 \xdef\sine{\Savesine}}}	      

\def\psdraft{
	\def\@psdraft{0}
}
\def\psfull{
	\def\@psdraft{100}
}

\psfull

\newif\if@scalefirst
\def\psscalefirst{\@scalefirsttrue}
\def\psrotatefirst{\@scalefirstfalse}
\psrotatefirst

\newif\if@draftbox
\def\psnodraftbox{
	\@draftboxfalse
}
\def\psdraftbox{
	\@draftboxtrue
}
\@draftboxtrue

\newif\if@prologfile
\newif\if@postlogfile
\def\pssilent{
	\@noisyfalse
}
\def\psnoisy{
	\@noisytrue
}
\psnoisy
\newif\if@bbllx
\newif\if@bblly
\newif\if@bburx
\newif\if@bbury
\newif\if@height
\newif\if@width
\newif\if@rheight
\newif\if@rwidth
\newif\if@angle
\newif\if@clip
\newif\if@verbose
\def\@p@@sclip#1{\@cliptrue}
\newif\if@decmpr
\def\@p@@sfigure#1{\def\@p@sfile{null}\def\@p@sbbfile{null}\@decmprfalse
   \openin1=\ps@predir#1
   \ifeof1
	\closein1
	\get@dir{#1}
	\ifx\ps@founddir\leer
		\openin1=\ps@predir#1.bb
		\ifeof1
			\closein1
			\get@dir{#1.bb}
			\ifx\ps@founddir\leer
				\ps@typeout{Can't find #1 in \figurepath}
			\else
				\@decmprtrue
				\def\@p@sfile{\ps@founddir\ps@dir#1}
				\def\@p@sbbfile{\ps@founddir\ps@dir#1.bb}
			\fi
		\else
			\closein1
			\@decmprtrue
			\def\@p@sfile{#1}
			\def\@p@sbbfile{#1.bb}
		\fi
	\else
		\def\@p@sfile{\ps@founddir\ps@dir#1}
		\def\@p@sbbfile{\ps@founddir\ps@dir#1}
	\fi
   \else
	\closein1
	\def\@p@sfile{#1}
	\def\@p@sbbfile{#1}
   \fi
}
\def\@p@@sfile#1{\@p@@sfigure{#1}}
\def\@p@@sbbllx#1{
		\@bbllxtrue
		\dimen100=#1
		\edef\@p@sbbllx{\number\dimen100}
}
\def\@p@@sbblly#1{
		\@bbllytrue
		\dimen100=#1
		\edef\@p@sbblly{\number\dimen100}
}
\def\@p@@sbburx#1{
		\@bburxtrue
		\dimen100=#1
		\edef\@p@sbburx{\number\dimen100}
}
\def\@p@@sbbury#1{
		\@bburytrue
		\dimen100=#1
		\edef\@p@sbbury{\number\dimen100}
}
\def\@p@@sheight#1{
		\@heighttrue
		\dimen100=#1
   		\edef\@p@sheight{\number\dimen100}
}
\def\@p@@swidth#1{
		\@widthtrue
		\dimen100=#1
		\edef\@p@swidth{\number\dimen100}
}
\def\@p@@srheight#1{
		\@rheighttrue
		\dimen100=#1
		\edef\@p@srheight{\number\dimen100}
}
\def\@p@@srwidth#1{
		\@rwidthtrue
		\dimen100=#1
		\edef\@p@srwidth{\number\dimen100}
}
\def\@p@@sangle#1{
		\@angletrue
		\edef\@p@sangle{#1} 
}
\def\@p@@ssilent#1{ 
		\@verbosefalse
}
\def\@p@@sprolog#1{\@prologfiletrue\def\@prologfileval{#1}}
\def\@p@@spostlog#1{\@postlogfiletrue\def\@postlogfileval{#1}}
\def\@cs@name#1{\csname #1\endcsname}
\def\@setparms#1=#2,{\@cs@name{@p@@s#1}{#2}}
\def\ps@init@parms{
		\@bbllxfalse \@bbllyfalse
		\@bburxfalse \@bburyfalse
		\@heightfalse \@widthfalse
		\@rheightfalse \@rwidthfalse
		\def\@p@sbbllx{}\def\@p@sbblly{}
		\def\@p@sbburx{}\def\@p@sbbury{}
		\def\@p@sheight{}\def\@p@swidth{}
		\def\@p@srheight{}\def\@p@srwidth{}
		\def\@p@sangle{0}
		\def\@p@sfile{} \def\@p@sbbfile{}
		\def\@p@scost{10}
		\def\@sc{}
		\@prologfilefalse
		\@postlogfilefalse
		\@clipfalse
		\if@noisy
			\@verbosetrue
		\else
			\@verbosefalse
		\fi
}
\def\parse@ps@parms#1{
	 	\@psdo\@psfiga:=#1\do
		   {\expandafter\@setparms\@psfiga,}}
\newif\ifno@bb
\def\bb@missing{
	\if@verbose{
		\ps@typeout{psfig: searching \@p@sbbfile \space  for bounding box}
	}\fi
	\no@bbtrue
	\epsf@getbb{\@p@sbbfile}
        \ifno@bb \else \bb@cull\epsf@llx\epsf@lly\epsf@urx\epsf@ury\fi
}	
\def\bb@cull#1#2#3#4{
	\dimen100=#1 bp\edef\@p@sbbllx{\number\dimen100}
	\dimen100=#2 bp\edef\@p@sbblly{\number\dimen100}
	\dimen100=#3 bp\edef\@p@sbburx{\number\dimen100}
	\dimen100=#4 bp\edef\@p@sbbury{\number\dimen100}
	\no@bbfalse
}
\newdimen\p@intvaluex
\newdimen\p@intvaluey
\def\rotate@#1#2{{\dimen0=#1 sp\dimen1=#2 sp
		  \global\p@intvaluex=\cosine\dimen0
		  \dimen3=\sine\dimen1
		  \global\advance\p@intvaluex by -\dimen3
		  \global\p@intvaluey=\sine\dimen0
		  \dimen3=\cosine\dimen1
		  \global\advance\p@intvaluey by \dimen3
		  }}
\def\compute@bb{
		\no@bbfalse
		\if@bbllx \else \no@bbtrue \fi
		\if@bblly \else \no@bbtrue \fi
		\if@bburx \else \no@bbtrue \fi
		\if@bbury \else \no@bbtrue \fi
		\ifno@bb \bb@missing \fi
		\ifno@bb \ps@typeout{FATAL ERROR: no bb supplied or found}
			\no-bb-error
		\fi
		\count203=\@p@sbburx
		\count204=\@p@sbbury
		\advance\count203 by -\@p@sbbllx
		\advance\count204 by -\@p@sbblly
		\edef\ps@bbw{\number\count203}
		\edef\ps@bbh{\number\count204}
		\if@angle 
			\Sine{\@p@sangle}\Cosine{\@p@sangle}
	        	{\dimen100=\maxdimen\xdef\r@p@sbbllx{\number\dimen100}
					    \xdef\r@p@sbblly{\number\dimen100}
			                    \xdef\r@p@sbburx{-\number\dimen100}
					    \xdef\r@p@sbbury{-\number\dimen100}}
                        \def\minmaxtest{
			   \ifnum\number\p@intvaluex<\r@p@sbbllx
			      \xdef\r@p@sbbllx{\number\p@intvaluex}\fi
			   \ifnum\number\p@intvaluex>\r@p@sbburx
			      \xdef\r@p@sbburx{\number\p@intvaluex}\fi
			   \ifnum\number\p@intvaluey<\r@p@sbblly
			      \xdef\r@p@sbblly{\number\p@intvaluey}\fi
			   \ifnum\number\p@intvaluey>\r@p@sbbury
			      \xdef\r@p@sbbury{\number\p@intvaluey}\fi
			   }
			\rotate@{\@p@sbbllx}{\@p@sbblly}
			\minmaxtest
			\rotate@{\@p@sbbllx}{\@p@sbbury}
			\minmaxtest
			\rotate@{\@p@sbburx}{\@p@sbblly}
			\minmaxtest
			\rotate@{\@p@sbburx}{\@p@sbbury}
			\minmaxtest
			\edef\@p@sbbllx{\r@p@sbbllx}\edef\@p@sbblly{\r@p@sbblly}
			\edef\@p@sbburx{\r@p@sbburx}\edef\@p@sbbury{\r@p@sbbury}
		\fi
		\count203=\@p@sbburx
		\count204=\@p@sbbury
		\advance\count203 by -\@p@sbbllx
		\advance\count204 by -\@p@sbblly
		\edef\@bbw{\number\count203}
		\edef\@bbh{\number\count204}
}
\def\in@hundreds#1#2#3{\count240=#2 \count241=#3
		     \count100=\count240	
		     \divide\count100 by \count241
		     \count101=\count100
		     \multiply\count101 by \count241
		     \advance\count240 by -\count101
		     \multiply\count240 by 10
		     \count101=\count240	
		     \divide\count101 by \count241
		     \count102=\count101
		     \multiply\count102 by \count241
		     \advance\count240 by -\count102
		     \multiply\count240 by 10
		     \count102=\count240	
		     \divide\count102 by \count241
		     \count200=#1\count205=0
		     \count201=\count200
			\multiply\count201 by \count100
		 	\advance\count205 by \count201
		     \count201=\count200
			\divide\count201 by 10
			\multiply\count201 by \count101
			\advance\count205 by \count201
		     \count201=\count200
			\divide\count201 by 100
			\multiply\count201 by \count102
			\advance\count205 by \count201
		     \edef\@result{\number\count205}
}
\def\compute@wfromh{
		\in@hundreds{\@p@sheight}{\@bbw}{\@bbh}
		\edef\@p@swidth{\@result}
}
\def\compute@hfromw{
	        \in@hundreds{\@p@swidth}{\@bbh}{\@bbw}
		\edef\@p@sheight{\@result}
}
\def\compute@handw{
		\if@height 
			\if@width
			\else
				\compute@wfromh
			\fi
		\else 
			\if@width
				\compute@hfromw
			\else
				\edef\@p@sheight{\@bbh}
				\edef\@p@swidth{\@bbw}
			\fi
		\fi
}
\def\compute@resv{
		\if@rheight \else \edef\@p@srheight{\@p@sheight} \fi
		\if@rwidth \else \edef\@p@srwidth{\@p@swidth} \fi
}
\def\compute@sizes{
	\compute@bb
	\if@scalefirst\if@angle
	\if@width
	   \in@hundreds{\@p@swidth}{\@bbw}{\ps@bbw}
	   \edef\@p@swidth{\@result}
	\fi
	\if@height
	   \in@hundreds{\@p@sheight}{\@bbh}{\ps@bbh}
	   \edef\@p@sheight{\@result}
	\fi
	\fi\fi
	\compute@handw
	\compute@resv}
\def\OzTeXSpecials{
	\special{empty.ps /@isp {true} def}
	\special{empty.ps \@p@swidth \space \@p@sheight \space
			\@p@sbbllx \space \@p@sbblly \space
			\@p@sbburx \space \@p@sbbury \space
			startTexFig \space }
	\if@clip{
		\if@verbose{
			\ps@typeout{(clip)}
		}\fi
		\special{empty.ps doclip \space }
	}\fi
	\if@angle{
		\if@verbose{
			\ps@typeout{(rotate)}
		}\fi
		\special {empty.ps \@p@sangle \space rotate \space} 
	}\fi
	\if@prologfile
	    \special{\@prologfileval \space } \fi
	\if@decmpr{
		\if@verbose{
			\ps@typeout{psfig: Compression not available
			in OzTeX version \space }
		}\fi
	}\else{
		\if@verbose{
			\ps@typeout{psfig: including \@p@sfile \space }
		}\fi
		\special{epsf=\ps@predir\@p@sfile \space }
	}\fi
	\if@postlogfile
	    \special{\@postlogfileval \space } \fi
	\special{empty.ps /@isp {false} def}
}
\def\DvipsSpecials{
	\special{ps::[begin] 	\@p@swidth \space \@p@sheight \space
			\@p@sbbllx \space \@p@sbblly \space
			\@p@sbburx \space \@p@sbbury \space
			startTexFig \space }
	\if@clip{
		\if@verbose{
			\ps@typeout{(clip)}
		}\fi
		\special{ps:: doclip \space }
	}\fi
	\if@angle
		\if@verbose{
			\ps@typeout{(clip)}
		}\fi
		\special {ps:: \@p@sangle \space rotate \space} 
	\fi
	\if@prologfile
	    \special{ps: plotfile \@prologfileval \space } \fi
	\if@decmpr{
		\if@verbose{
			\ps@typeout{psfig: including \@p@sfile.Z \space }
		}\fi
		\special{ps: plotfile "`zcat \@p@sfile.Z" \space }
	}\else{
		\if@verbose{
			\ps@typeout{psfig: including \@p@sfile \space }
		}\fi
		\special{ps: plotfile \@p@sfile \space }
	}\fi
	\if@postlogfile
	    \special{ps: plotfile \@postlogfileval \space } \fi
	\special{ps::[end] endTexFig \space }
}
\def\psfig#1{\vbox {
	\ps@init@parms
	\parse@ps@parms{#1}
	\compute@sizes
	\ifnum\@p@scost<\@psdraft{
		\PsfigSpecials 
		\vbox to \@p@srheight sp{
			\hbox to \@p@srwidth sp{
				\hss
			}
		\vss
		}
	}\else{
		\if@draftbox{		
			\hbox{\fbox{\vbox to \@p@srheight sp{
			\vss
			\hbox to \@p@srwidth sp{ \hss 
			 \hss }
			\vss
			}}}
		}\else{
			\vbox to \@p@srheight sp{
			\vss
			\hbox to \@p@srwidth sp{\hss}
			\vss
			}
		}\fi

	}\fi
}}
\psfigRestoreAt
\setDriver
\let\@=\LaTeXAtSign

\documentstyle{cupconf}
\setcounter{page}{1}

\newcommand\etal{{ et al. }}
\def\mdot{$\dot M$}
\def\msy{$M_\odot$ yr$^{-1}$}
\def\kms{km s$^{-1}$}
\def\e#1{$\times$ $10^{#1}$ }
\def\ee#1{$10^{#1}$ }
\def\ms{$M_\odot$}
\def\ni{$^{56}$Ni}
\def\co{$^{56}$Co}
\def\fe{$^{56}$Fe}
\def\ri{$R_{\rm i}$}
\def\slum{$L_\odot$}
\def\srad{$R_\odot$}
\def\gam{$\gamma$}
\def\lsim{\mathrel{\rlap{\lower 4pt \hbox{\hskip 1pt $\sim$}}\raise 1pt \hbox
        {$<$}}}
\def\gsim{\mathrel{\rlap{\lower 4pt \hbox{\hskip 1pt $\sim$}}\raise 1pt \hbox
        {$>$}}}
\def\ltsim{\mathrel{\rlap{\lower 4pt \hbox{\hskip 1pt $\sim$}}\raise 1pt \hbox
        {$<$}}}
\def\gtsim{\mathrel{\rlap{\lower 4pt \hbox{\hskip 1pt $\sim$}}\raise 1pt \hbox
        {$>$}}}
\def\apj{ApJ}
\def\aap{A\&A}
\def\aj{AJ}
\def\mnras{MNRAS}
\def\araa{ARAA}
\def\pasp{PASP}
\def\apjs{ApJS}
\def\nat{Nature}

\newcommand{\eg}{e.g.\ }
\newcommand{\ie}{i.e.\ }
\newcommand{\Msun}{M_{\odot}}
\newcommand{\Rsun}{R_{\odot}}
\newcommand{\Lsun}{L_{\odot}}
\newcommand{\ergs}{erg s$^{-1}$}
\newcommand{\Ha}{H$\alpha$} 
\newcommand{\Hb}{H$\beta$}
\newcommand{\HI}{H~{\sc i}}
\newcommand{\OI}{O~{\sc i}}
\newcommand{\NaI}{Na~{\sc i}}
\newcommand{\MgII}{Mg~{\sc ii}}
\newcommand{\MgI}{Mg~{\sc i}}
\newcommand{\CaII}{Ca~{\sc ii}}
\newcommand{\SiI}{Si~{\sc i}}
\newcommand{\SiII}{Si~{\sc ii}}
\newcommand{\SiIII}{Si~{\sc iii}}
\newcommand{\TiII}{Ti~{\sc ii}}
\newcommand{\CrII}{Cr~{\sc ii}}
\newcommand{\FeI}{Fe~{\sc i}}
\newcommand{\FeII}{Fe~{\sc ii}}
\newcommand{\FeIII}{Fe~{\sc iii}}
\newcommand{\CoII}{Co~{\sc ii}}
\newcommand{\NiII}{Ni~{\sc ii}}
\newcommand{\Fefs}{$^{56}$Fe}
\newcommand{\Cofs}{$^{56}$Co}
\newcommand{\Nifs}{$^{56}$Ni}
\newcommand{\Mej}{$M_{\rm ej}$}
\newcommand{\KE}{$E_{\rm K}$}

\title[Hypernovae]
{
{\small {\it To be published in "Supernovae and Gamma Ray Bursts" \\
\vspace{-.3cm}
ed. M. Livio et al. (Cambridge University Press)}}\\
\vspace{2mm}
The Properties of Hypernovae: \\
SNe Ic 1998bw, 1997ef, and SN IIn 1997cy
}

\author[Nomoto et al.]
{K. NOMOTO$^{1,2}$, P.A. MAZZALI$^{2,3}$, T. NAKAMURA$^1$,\\
K. IWAMOTO$^4$, K. MAEDA$^1$, T. SUZUKI$^{1,2}$, \\
M. TURATTO$^{5}$, I.J. DANZIGER$^{3}$, F. PATAT$^{6}$}
\affiliation{$^1$Department of Astronomy, School of Science, 
University of Tokyo, Tokyo, Japan\\
$^2$Research Center for the Early Universe, School of Science, 
University of Tokyo, Tokyo, Japan\\
$^3$Osservatorio Astronomico di Trieste, via G. B. Tiepolo, Trieste, Italy\\
$^4$Department of Physics, College of Science and Technology,
Nihon University, Tokyo, Japan\\
$^5$Osservatorio Astronomico di Padova, vicolo dell'Osservatorio, Padova,
Italy\\
$^6$European Southern Observatory, Garching, Germany
}

\begin{document}

\maketitle


\begin{abstract}

We discuss the properties of the hyper-energetic Type Ic supernovae
(SNe Ic) 1998bw and 1997ef and Type IIn supernova (SN IIn) 1997cy.
SNe Ic 1998bw and 1997ef are characterized by their large luminosity
and the very broad spectral features.  Their observed properties can
be explained if they are very energetic SN explosions with the kinetic
energy of $E_{\rm K} \gsim 1\times10^{52}$ erg, originating probably
from the core collapse of the bare C+O cores of massive stars ($\sim
30-40$M$_\odot$).  At late times, both the light curves and the
spectra suggest that the explosions may have been asymmetric; this may
help us understand the claimed connection with GRB's.  The Type IIn SN
1997cy is even more luminous than SN 1998bw and the light curve
declines more slowly than $^{56}$Co decay.  We model such a light
curve with circumstellar interaction, which requires the explosion
energy of $\sim 3 \times 10^{52}$ erg. Because these kinetic energies
of explosion are much larger than in normal core-collapse SNe, we call
objects like these SNe "hypernovae".  The mass of \ni\ in SN 1998bw is
estimated to be as large as 0.5 - 0.7 \ms\ from both the maximum
brightness and late time emission spectra, which suggests that the
asymmetry may not be extreme.

\end{abstract}

\section{Introduction}

Recently, there have been an increasing number of candidates for the
gamma-ray burst (GRB)/supernova (SN) connection (Woosley 1993;
Paczy\'nski 1998).  The first example of such a candidate was provided
by SN~1998bw.  SN~1998bw was discovered in the error box of GRB980425
(Kulkarni et al. 1998), only 0.9 days after the date of the gamma-ray
burst and was very possibly linked to it (Galama \etal 1998).

\begin{figure}
\centerline{\psfig{figure=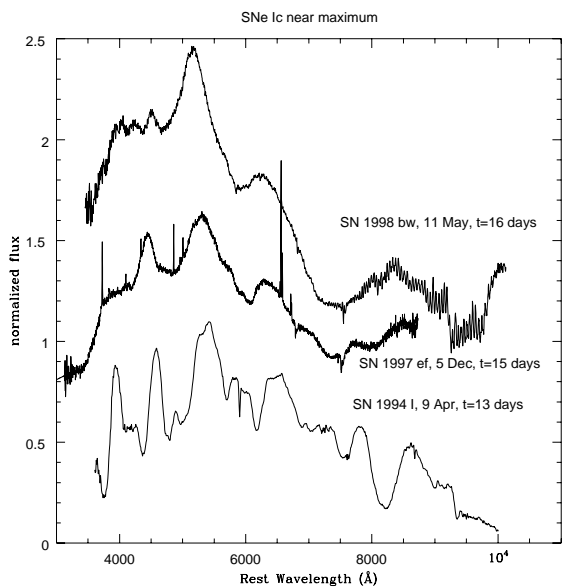,width=8cm}}
\caption{Observed spectra of Type Ic supernovae 1998bw, 1997ef, and 1994I.
\label{fig:spic}}
\end{figure}

\begin{figure}
\centerline{\psfig{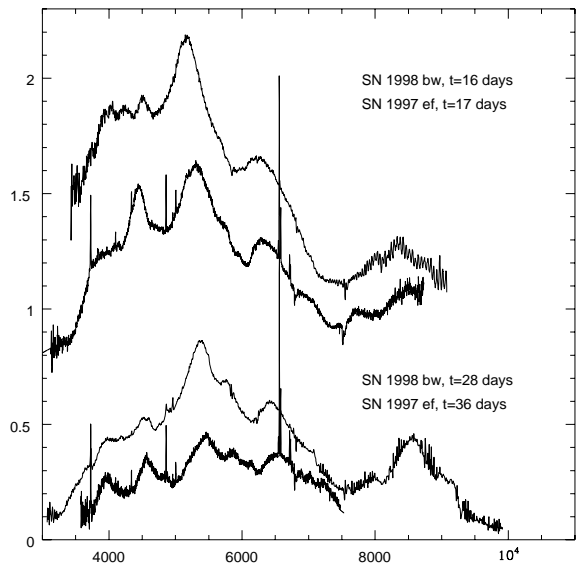}}
\caption{Observed spectra of Type Ic supernovae 1998bw and 1997ef.
\label{fig:spic2}}
\end{figure}

\begin{figure}
\vspace{1cm}
\centerline{\psfig{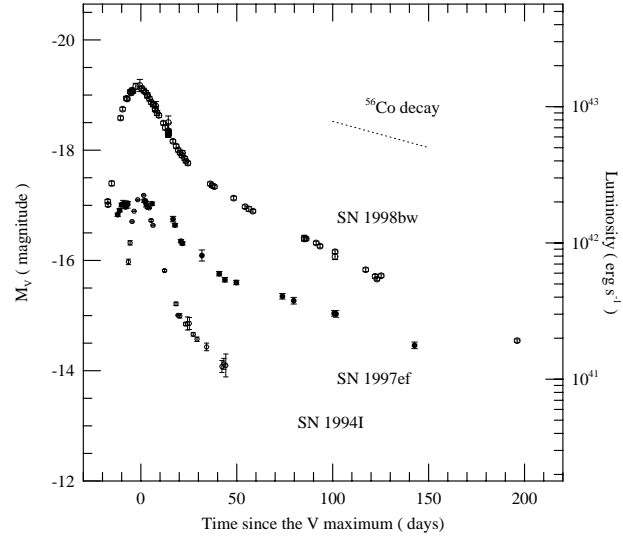}}
\caption{Absolute magnitudes of Type Ic supernovae: the ordinary SN~Ic
1994I (Richmond \etal 1996a, b), and the hypernovae SN~1998bw (Galama et
al. 1998) and SN~1997ef (Iwamoto et al. 2000).  The dashed line
indicates the \(^{56}\)Co decay rate.  
\label{fig:lcic}}
\end{figure}

\begin{figure}
\centerline{\psfig{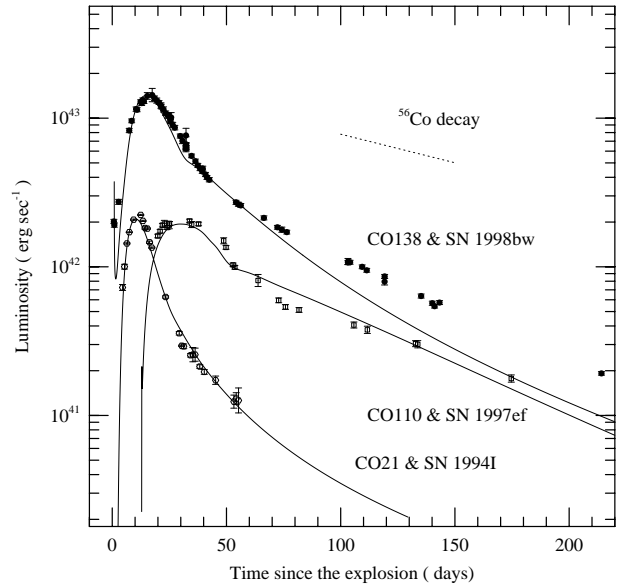}}
\caption{
Same as Figure \ref{fig:lcic} but with theoretical models.
\label{fig:lcic2}}
\end{figure}

Early spectra were rather blue and featureless, showing some
similarities with the spectra of Type Ic SNe (SNe Ic),
but with one major difference
(Figs. \ref{fig:spic}, \ref{fig:spic2}):
the absorption lines were so broad in SN~1998bw that they
blended together, giving rise to broad absorption trough separated by
apparent 'emission peaks' (Iwamoto et al. 1998; Patat \etal 2000;
Stathakis et al. 2000).
This supernova was immediately recognized to be very powerful and
bright (Figs. \ref{fig:lcic}, \ref{fig:lcic2}).

Velocities in the Si~{\sc ii} 6355\AA\ line are as high as $30,000$
km~s$^{-1}$. Also, the SN was very bright for a SN~Ic: the observed
peak luminosity, $L \sim 1.4 \times 10^{43}$ erg s$^{-1}$, is almost
ten times higher than that of previously known SNe Ib/Ic.
Models which described the SN as the energetic explosion of a
C+O core of an initially massive star could successfully fit the first
60 days of the light curve (Iwamoto et al. 1998; hereafter IMN98).

The very broad spectral features and the light curve shape have led to
the conclusion that SN 1998bw had an extremely large {\sl kinetic}
energy of explosion, $E_{\rm K} \sim$ 3 \e{52} ergs (IMN98; Woosley,
Eastman, \& Schmidt 1999; Branch 2000).
This is more than one order of magnitude
larger than the energy of typical supernovae, thus SN 1998bw was
termed a ``hypernova'' (IMN98).  ``Hypernova'' is a term we use to describe
the events of $E_{\rm K} \gsim 10^{52}$ erg without specifying whether
the central engine is a collapsar (MacFadyen \& Woosley 1999)
or magnetar (Wheeler et al. 2000) or pair-instability
(Woosley 2000; Nakatsuru et al. 2000).

SN~1997ef was also noticed for its unique light curve and spectra
(Figs. \ref{fig:spic} - \ref{fig:lcic2}).  At early times, the spectra
were dominated by broad oxygen and iron absorption lines, but did not
show any clear feature of hydrogen or helium (Garnavich et al. 1997a;
Hu et al. 1997), which led us to classify SN~1997ef as a SN Ic.  The
most striking and peculiar characteristic of SN~1997ef is the width of
its line features.  Such broad spectral features were later recognized
to be a distinguishing property of the spectra in SN 1998bw
(Figs. \ref{fig:spic}, \ref{fig:spic2}).  The spectral similarities
between SN 1997ef and SN 1998bw suggest that SN 1997ef may also be a
hypernova (Iwamoto et al. 2000; Nomoto et al. 1999; Branch 2000).

In Figure \ref{fig:lcic} the visual light curve of SNe 1998bw (Galama
et al. 1998) and 1997ef (Garnavich \etal 1997b, c) are compared with
the ordinary SN~Ic 1994I (Richmond \etal 1996a, b).  Despite the
spectral similarity, the light curve of SN~1997ef is quite different
from those of SN~1998bw and SN~1994I.  It has a flat peak, much
broader than those of the other SNe Ic.  Besides, the tail of the
light curve of SN~1997ef starts late and the rate of its decline is
much slower than in other SNe Ic.  Since the light curves are rather
diverse, even in this limited number of samples, a range of energies
and/or progenitor masses of SN Ic explosions may be implied.

Among the SNe with a possible GRB counterpart, we show here that the
Type IIn SN 1997cy (Germany et al. 1999; Turatto et al. 1999, 2000) is also
characterized by an extremely large kinetic explosion energy, $E_{\rm
K} \gsim 10^{52}$ erg, thus belonging to "hypernovae".  Also, SN IIn
1999E has a spectrum very similar to that of SN~1997cy (Cappellaro et
al. 1999), and so it is probably a similar object.

Despite the success of our 'hypernova' model in reproducing the early
light curves of SN~1998bw and SN~1997ef, the model light curve tail
declines more rapidly than the observations for both SNe.  Another
unexpected feature appeared in the late phase spectra of SN1998bw.
Measuring the velocity of each element (Patat et al. 2000), we note
that iron expands faster than oxygen, which is contrary to
expectations.  This may indicate asymmetry in the ejecta.

Here we summarize the photometric and spectroscopic properties of
these hypernovae and the estimated explosion energies and ejecta mass
using the hydrodynamical models.  We also study nucleosynthesis in
hypernovae in 1D and 2D.

\section{Explosion Models for Supernovae and Hypernovae}\label{sec:model}

We construct hydrodynamical models of an ordinary SN Ic and a
hypernova as follows.  Since the light curve of SN Ic 1994I was
successfully reproduced by the collapse-induced explosion of C+O stars
(Nomoto et al. 1994; Iwamoto et al. 1994), we adopted C+O stars as
progenitor models for SNe 1997ef and 1998bw as well.  We calculate the
light curves and spectra for various C+O star models with different
values of $E_{\rm K}$ and $M_{\rm ej}$.  These parameters can be
constrained by comparing the calculated light curves, the synthetic
spectra, and the photospheric velocities with the data of SNe 1998bw
and 1997ef.

(1) In the ordinary SN Ic model (model CO60), a C+O star with a mass
$M_{\rm CO}= 6.0 M_\odot$ (which is the core of a 25 $M_\odot$
main-sequence star; Nomoto \& Hashimoto 1988) explodes with kinetic
energy of explosion $E_{\rm K} = 1.0 \times 10^{51}$ ergs and ejecta
mass $M_{\rm ej} = M_{\rm CO}-M_{\rm cut} = 4.6 M_\odot$. Here $M_{\rm
cut}$ (= 1.4 \ms) denotes the mass of the compact star remnant (either
a neutron star or a black hole).

(2) In the hypernova model (CO100), a C+O star of $M_{\rm CO}= 10
M_\odot$ is constructed from a 10 \ms\ He star (which has a 8 \ms\ C+O
core) by removing the outermost 2 \ms\ of He layer and extending the
C+O layer up to $10 M_\odot$.  This model corresponds to 30 - 35
$M_\odot$ on the main-sequence.  The star explodes with $E_{\rm K} =
8.0 \times 10^{51}$ ergs, ejecting $M_{\rm ej} = 7.6 M_\odot$, i.e.,
$M_{\rm cut}$ = 2.4 \ms.

(3) For the hypernova models CO138H and CO138L,
C+O stars of $M_{\rm CO}= 13.8
M_\odot$ are constructed from a 16 \ms\ He star.
This model corresponds to $\sim$ 40 $M_\odot$ on the main-sequence.
These models are exploded with
$E_{\rm K} = 6 \times 10^{52}$ erg (CO138H) and
$E_{\rm K} = 3 \times 10^{52}$ erg (CO138L) and
$M_{\rm ej} \simeq 11M_\odot$, i.e., $M_{\rm cut} \simeq$ 3 - 4 \ms.

These model parameters are summarized in Table 1, together with model CO21
for SN 1994I.  The position of the mass cut is chosen so that the ejected
mass of $^{56}$Ni is the value required to explain the observed peak
brightness
of SN~1997ef and SN~1998bw by radioactive decay heating.  The compact
remnant in CO60 is probably a neutron star because $M_{\rm cut}$ = 1.4
\ms, while it may be a black hole in CO100 and CO138 because $M_{\rm
cut}$ may well exceed the maximum mass of a stable neutron star.

The hydrodynamics at early phases was calculated by using a Lagrangian
PPM code (Colella \& Woodward 1984).  The explosion is triggered by
depositing thermal energy in a couple of zones just below the mass cut
so that the final kinetic energy has the required value.  The
explosive nucleosynthesis are discussed in \S \ref{sec:nuc}.

\begin{figure}
\centerline{\psfig{figure=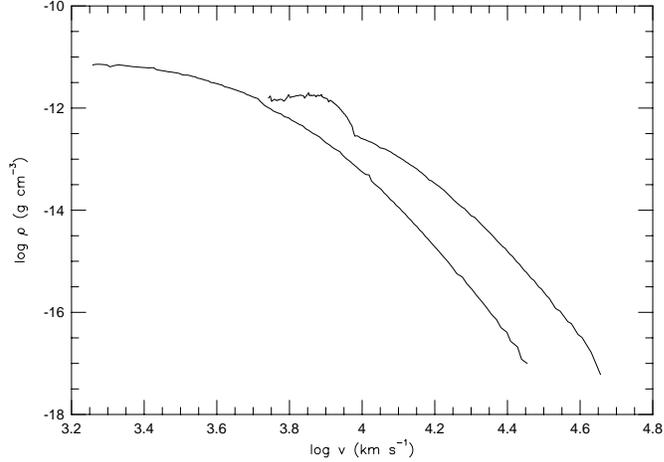,width=10.cm}}
\caption{Density distributions against the velocity of
homologously expanding ejecta for CO60 and CO100.
\label{fig:vrhonew}}
\end{figure}

The expansion soon becomes homologous so that $v \propto r$.  The
solid lines in Figure \ref{fig:vrhonew} show the density distributions
in velocity space for CO60 and CO100 at $t =$ 16 days.  The expansion
velocities are clearly higher in CO100 than in CO60.  We performed
detailed radiation transfer calculations to obtain light curves and
spectra for the explosion models.  The results were compared with
observations of SNe 1998bw and 1997ef, to derive explosion energies
and the ejecta masses, and thus to determine whether the SNe were
ordinary SNe Ic or hypernovae.

\begin{table}
\begin{center}
\centerline{Table 1.~Parameters of the C+O star models}
\vspace*{4mm}
\begin{tabular}{cccccccccc}
\hline model & $M_{\rm ms}${\small($M_\odot$)}
& $M_{\rm C+O}$ & $M_{\rm ej}$ &
 $^{56}$Ni mass & $M_{\rm cut}$
& $E_{\rm K}$ {\small (10$^{51}$ erg)}& SN \\
\hline CO21  & $\sim$ 15 &  2.1 &  0.9 & 0.07 & 1.2  & 1 & 1994I \\ 
\hline CO60  & $\sim$ 25 &  6.0 &  4.4 & 0.15 & 1.4 & 1 & \\ 
\hline CO100 & $\sim$ 30 - 35 & 10.0 &  7.6 & 0.15 & 2.4 & 8 & 1997ef   \\ 
\hline CO138H & $\sim$ 40 & 13.8 & 10 & 0.5 &  4 & 60 & 1998bw \\ 
\hline CO138L & $\sim$ 40 & 13.8 & 11 & 0.5 & 3 & 30 & 1998bw  \\ 
\hline
\end{tabular}
\end{center}
\end{table}

\section{SN~1997ef}\label{sec:97ef}

\begin{figure}
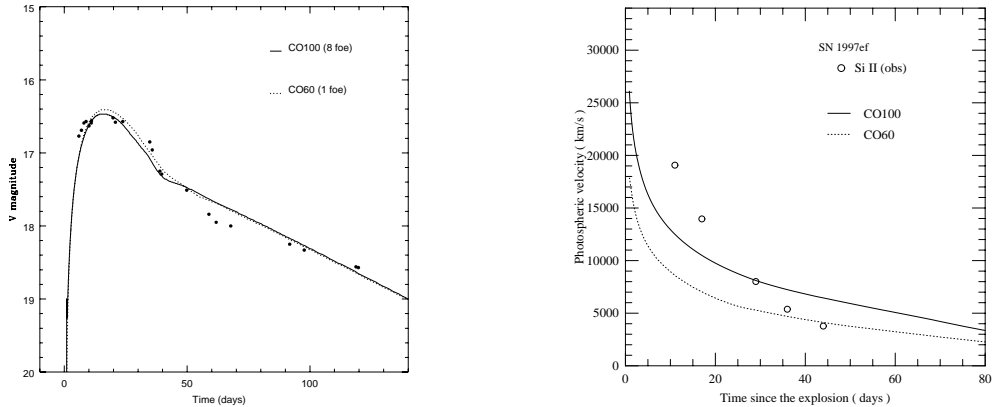

\vspace*{2.cm}
\hspace*{-1cm}
\begin{center}\leavevmode
\psfig{figure=lc97ef.epsi,width=5.3cm}
\hspace*{2.cm}
\psfig{figure=vph97ef.epsi,width=5.5cm}
\end{center}
\caption{
Left panel: Calculated Visual light curves of CO60 and CO100 compared
with that of SN 1997ef.
Right panel: Evolution of the calculated photospheric velocities of
CO60 and CO100 (solid lines) compared with the observed velocities of the 
Si II 634.7, 637.1 nm line measured in the spectra at the absorption core.
\label{fig:lc97ef}}
\end{figure}



\subsection{Light Curve Models}

In Figure \ref{fig:lc97ef} we compare the calculated V light curves for
models CO60 and CO100 with the observed V light curve of SN1997ef.  We
adopt a distance of 52.3 Mpc (a distance modulus of $\mu=33.6$ mag) as
estimated from the recession velocity, 3,400 km s$^{-1}$ (Garnavich
\etal 1997a) and a Hubble constant $H_0=65$ km s$^{-1}$ Mpc$^{-1}$.  We
assume no color excess, $E(B-V)=0.00$; this is justified by the fact
that no signature of a narrow Na I D interstellar absorption line is
visible in the spectra of SN~1997ef at any epochs (Garnavich et al. 1997a).
The light curve of SN 1997ef has a very broad maximum, which lasts for
$\sim$ 25 days.  The light curve tail starts only $\sim 40$ days after
maximum, much later than in other SNe Ic.

The light curve of SN 1997ef can be reproduced basically with various
explosion models with different energies and masses.  In general, the
properties of the light curve are characterized by the decline rate in
the tail and the peak width, $\tau_{\rm peak}$.  The peak width scales
approximately as
\begin{equation}
\tau_{\rm peak} \propto \kappa^{1/2} M_{\rm ej}^{3/4} E_{\rm
K}^{-1/4},
\end{equation}
where $\kappa$ denotes the optical opacity (Arnett 1996).  This is the
time-scale on which photon diffusion and hydrodynamical expansion
become comparable.  Since the model parameters of CO100 and CO60 give
similar $\tau_{\rm peak}$, the light curves of the two models look
similar: both have quite a broad peak and reproduce the light curve of
SN1997ef reasonably well (Figure \ref{fig:lc97ef}).

The light curve shape depends also on the distribution of $^{56}$Ni,
which is produced in the deepest layers of the ejecta. More extensive
mixing of \ni\ leads to an earlier rise of the light curve.  For SN
1997ef, the best fit is obtained when $^{56}$Ni is mixed almost
uniformly to the surface for both models.  Without such extensive
mixing, the rise time to V $=$ 16.5 mag would be $\sim$ 30 d for
CO100, which is clearly too long to be compatible with the
spectroscopic dating.

Model CO60 has the same kinetic energy ($E_{\rm K} = 1$ \e{51} erg) as
model CO21, which was used for SN Ic 1994I (see Table 1 for the model
parameters). Since the light curve of SN 1997ef is much slower than
that of SN 1994I, the ejecta mass of CO60 is $\sim$ 5 times larger
than that of CO21.

The ejecta mass of CO100 is a factor of $\sim 2$ larger than that of
CO60, and it is only $\sim 20$\% smaller than that of model CO138,
which was used for SN 1998bw (Table 1).  Thus the explosion energy of
CO100 should be $\sim 8$ times larger than that of CO60 to reproduce
the light curve of SN 1997ef. This explosion is very energetic, but
still much weaker than the one in CO138.  The smaller $E_{\rm K}$ for
a comparable mass allows CO100 to reproduce the light curve of SN
1997ef, which has a much broader peak than that of SN 1998bw.

The light curve of SN 1997ef enters the tail around day 40.  Since
then, the observed V magnitude declines linearly with time at a rate
of $\sim 1.1 \times 10^{-2}$ mag day$^{-1}$, which is slower than in
other SNe Ic and is close to the $^{56}$Co decay rate $9.6 \times
10^{-3}$ mag day$^{-1}$.  Such a slow decline implies much more
efficient $\gamma$-ray trapping in the ejecta of SN 1997ef than in SN
1994I.  The ejecta of both CO100 and CO60 are fairly massive and are
able to trap a large fraction of the $\gamma$-rays, so that the
calculated light curves have slower tails compared with CO21.

However, the light curves for both models decline somewhat faster in
the tail than the observations.  A similar discrepancy has been noted
for the Type Ib supernovae (SNe Ib) 1984L and 1985F (Swartz \& Wheeler
1991; Baron et al. 1993).  The late time light curve decline of these
SNe Ib is as slow as the \co\ decay rate, so that the inferred value
of $M$ is significantly larger (and/or $E_{\rm K}$ is smaller) than
those obtained by fitting the early light curve shape.  Baron et
al. (1993) suggested that the ejecta of these SNe Ib must be highly
energetic and as massive as $\sim$ 50 \ms.  In \S \ref{sec:98bwlc}, we
will suggest that such a discrepancy between the early- and late-time
light curves might be an indication of asphericity in the ejecta of SN
1997ef and that it might be the case in those SNe Ib as well.

\subsection{Photospheric Velocities}\label{sec:97efvph}

As we have shown, light curve modeling provides direct constraints on
$M_{\rm CO}$ and $E_{\rm K}$.  However, it is difficult to distinguish
between the ordinary SN Ic and the hypernova model from the light
curve shape alone, since models with different values of $M_{\rm ej}$
and $E_{\rm K}$ can reproduce similar light curves.  However, these
models are expected to show different evolutions of the photospheric
velocity and the spectrum as will be discussed in the following
sections.

The photospheric velocity scales roughly as $v_{\rm ph} \propto M_{\rm
ej}^{-1/2} E_{\rm K}^{1/2}$, so that $M_{\rm ej}$ and $E_{\rm K}$ can
be constrained by $v_{\rm ph}$ in a different way from by means of the
light curve width.  Figure \ref{fig:lc97ef} (right) shows the
evolution of the observed velocities of the Si~II line measured in the
spectra at the absorption core, and the velocities at the grey
photosphere computed by the light curve code for models CO60 and
CO100.  The velocities of the Si~II line are somewhat higher than that
of the photosphere, reaching $\sim$ 20,000 \kms at the earliest times.

In model CO60 the photosphere forms at velocities much smaller than
those of the observed lines, while CO100 gives photospheric velocities
as high as the observed ones.  It is clear, from this comparison, that
the hyper-energetic model CO100 is preferable to the ordinary model
CO60.  The apparent discrepancy that still exists between the model
CO100 and observations might be related to the morphology of the
ejecta, i.e., a deviation from spherical symmetry, as was also
suggested in the case of SN~1998bw (H\"oflich et al. 1999; IMN98).

\subsection{Synthetic Spectra} 



To strengthen the arguments in \S \ref{sec:97efvph}, we compare the
emergent spectra for the two explosion models.  Using detailed spectrum
synthesis, we can distinguish between different models more clearly,
because the spectrum contains much more information than a single-band
light curve.

\begin{figure}[hu]
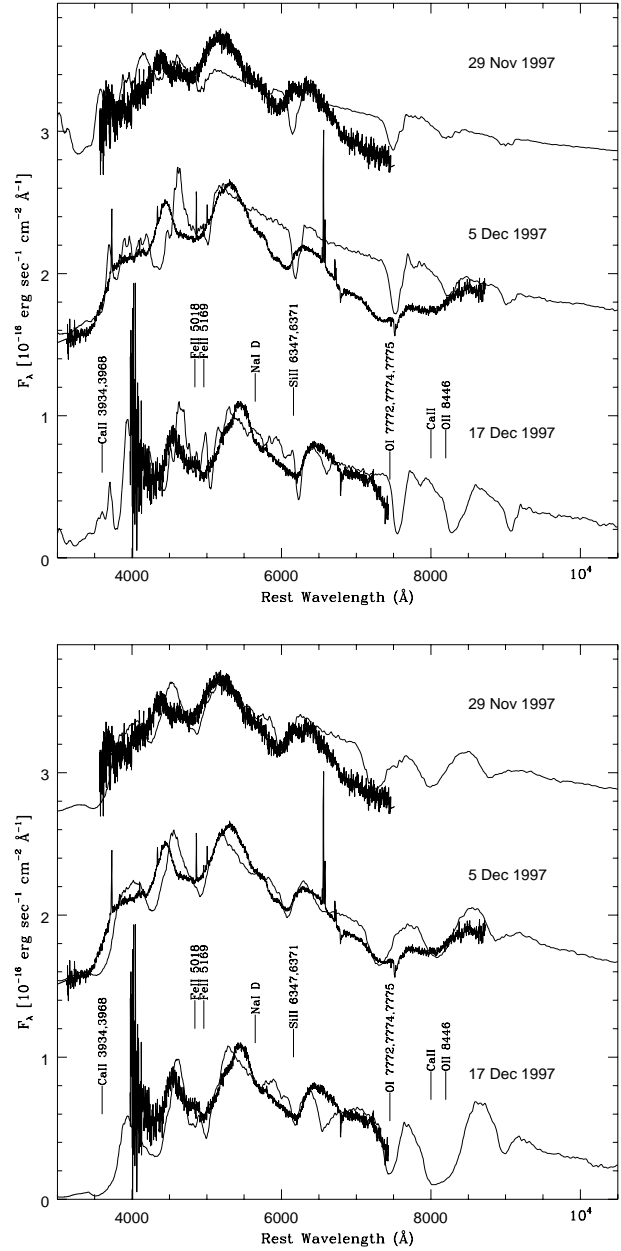

\centerline{\psfig{figure=97efspc1.epsi,width=8.cm}}
\vspace{.5cm}
\centerline{\psfig{figure=97efspc2.epsi,width=8.cm}}
\caption{
Upper panel: Observed spectra of SN~1997ef (bold lines)
and synthetic spectra computed 
using model CO60. The lines in the synthetic spectra are much too narrow.
Lower panel: Observed spectra of SN~1997ef (bold lines)
and synthetic spectra computed using model CO100 (fully drawn lines).
\label{fig:97efspc}}
\end{figure}

Around maximum light, the spectra of SN~1997ef show just a few very
broad features, and are quite different from those of ordinary SNe
Ib/c, but similar to SN~1998bw.  However, at later epochs the spectra
develop features that are easy to identify, such as the Ca~II IR
triplet at $\sim 8200$\AA, the O~I absorption at 7500 \AA, several
Fe~II features in the blue, and they look very similar to the spectrum
of the ordinary SN Ic 1994I.

We computed synthetic spectra with a Monte Carlo spectrum synthesis
code using the density structure and composition of the hydrodynamic
models CO60 and CO100.  The code is based on the pure scattering code
described by Mazzali \& Lucy (1993), but has been improved to include
photon branching, so that the reprocessing of the radiation from the
blue to the red is followed more accurately and efficiently (Lucy
1999; Mazzali 2000).

We produced synthetic spectra for three epochs near maximum, of
SN~1997ef: Nov 29, Dec 5, and Dec 17. These are early enough that the
spectra are very sensitive to changes in the kinetic energy.  As in
the light curve comparison, we adopted a distance modulus of
$\mu=33.6$ mag, and $E(B-V)=0.0$.

In Figure \ref{fig:97efspc} (above) we show the synthetic spectra
computed with the ordinary SN~Ic model CO60.  The lines in the spectra
computed with this model are always much narrower than the observations.
This clearly indicates a lack of material at high velocity in model
CO60, and suggests that the kinetic energy of this model is much too
small.

Synthetic spectra obtained with the hypernova model CO100 for the same 3
epochs are shown in Figure \ref{fig:97efspc} (below).  The spectra show
much broader lines, and are in good agreement with the observations.  In
particular, the blending of the Fe lines in the blue, giving rise to
broad absorption troughs, is well reproduced, and so is the very broad
Ca-O feature in the red.  The two `emission peaks' observed at $\sim
4400$ and 5200\AA\ correspond to the only two regions in the blue that
are relatively line-free.

The spectra are characterized by a low temperature, even near maximum,
because the rapid expansion combined with the relatively low
luminosity (from the tail of the light curve we deduce that SN~1997ef
produced about $0.15 M_\odot$ of $^{56}$Ni, compared to about $0.6
M_\odot$ in a typical SN~Ia and $0.5 M_\odot$ in SN~1998bw) leads to
rapid cooling.  Thus the \SiII\ 6355\AA\ line is not very strong.

Although model CO100 yields rather good synthetic spectra, it still
fails to reproduce the observed large width of the O~I - Ca~II feature
in the only near-maximum spectrum that extends sufficiently far to the
red (5 Dec 1997). An improvement can be obtained by introducing an
arbitrary flattening of the density profile at the highest velocities
(Mazzali, Iwamoto \& Nomoto 2000; Branch 2000).

\subsection{Possible Aspherical Effects}

We have shown that the light curve, the photospheric velocities, and
the spectra of SN~1997ef are better reproduced with the
hyper-energetic model CO100 than with the ordinary SN Ic model
CO60. However, there remain several features that are still difficult
to explain with model CO100.

(1) The observed velocity of Si II decreases much more rapidly than
models predict. It is as high as $\sim$ 30,000 \kms\ at the earliest
phase, but it gets as low as $\sim$ 3,000 \kms\ around day 50 (Figure
\ref{fig:lc97ef}, right).  We find that it is difficult to get such a
rapid drop of the photospheric velocity not only in models CO100 and
CO60, but also in other models that can reproduce the light-curve
shape reasonably well.  Models with higher energies and/or smaller
masses would be able to reproduce the fast evolution of the
photospheric velocity, but such models would inevitably produce light
curves with a narrower peak and a faster tail.

(2) Obviously, the observed light curve decline is slower than model
CO100 in the tail part, and it is also a bit flatter than the model
around maximum (Figure \ref{fig:lc97ef}).  Models with lower energies
and/or larger masses give better fits to both the peak and the tail of
the light curve, but then it gets very difficult to reproduce the large
photospheric velocities observed at early times in SN~1997ef.

This dilemma might be overcome if we introduce multiple components of
the light curve from different parts of ejecta moving at different
velocities. In fact, the discrepancies may be interpreted as a
possible sign of asphericity in the ejecta: A part of the ejecta moves
faster than average to form the lines at high-velocities at early
phases, while the other part of ejecta expands with a lower velocity
so that the low-velocity \SiII\ line comes up at later epochs. Having a
low-velocity component would also make it easier to reproduce the slow
tail.

(3) Extensive mixing of $^{56}$Ni is required to reproduce the short
rise time of the light curve.  According to hydrodynamical simulations
of the Rayleigh-Taylor instability in the ejecta of envelope-stripped
supernovae (Hachisu \etal 1991; Iwamoto \etal 1997), large scale
mixing is not expected to occur in massive progenitors, because in the
core of such massive stars the density gradient is not steep enough
around the composition interfaces.  One possibility to induce such
mixing in the velocity space is an asymmetric explosion.  Higher
velocity \ni\ could reach the ejecta surface so that the effect of
radioactive heating comes up as early as is required from light curve
modeling.

In order to realize higher densities at low velocity regions without
increasing the mass of the ejecta significantly, it may be necessary
that the explosion is somewhat aspherical.  If the explosion is
aspherical, the shock would be stronger and the material would expand
at a larger velocity in a certain direction, while in the
perpendicular direction the shock would be weaker, ejecting lower
velocity material (e.g., H\"oflich et al. 1999).  The density of
the central region could be high enough for $\gamma$-rays to be
trapped even at advanced phases, thus giving rise to a slowly
declining tail.  In the extremely asymmetric cases, material ejection
may take place in a jet-like form (e.g., MacFadyen \& Woosley 1999;
Khoklov et al. 1999).
A jet could easily bring some
\ni\ from the deepest layers out to the high velocity surface.
Detailed spectral analysis of observed spectra for different epochs is
necessary to investigate this issue further.

\section{SN~1998bw}\label{sec:98bw}

In this section models are presented for SN~1998bw which reproduce
most of the early data.  The very bright and relatively broad light
curve of SN~1998bw can be reproduced by a family of models with
various values of the fundamental parameters (\Mej, \KE), but in all
the models these parameters are much larger than in the case of a
typical SN~Ic.  All models require $M$(\Nifs)$ \sim 0.5 \Msun$ to
power the bright light curve peak. This is about an order of magnitude
larger than in typical core-collapse SNe. Models with different \KE\
yield different synthetic spectra, and by comparing with the observed
early-time spectra of SN~1998bw and trying to fit the very broad
absorption features, we selected a model CO138H with \Mej$ = 10
\Msun$, \KE$ = 6 \times 10^{52}$~erg (Nakamura et al. 1999b). The
large value of \KE\ easily qualifies SN~1998bw as `the' Type Ic
Hypernova.  The mass of the progenitor C+O star is $13.8 \Msun$, which
implies a main sequence mass of $\sim 40 \Msun$.  We also find that
imposing a flatter density structure at high velocities ($v >
30,000$~\kms) results in more high-velocity absorption and in an even
better-looking spectrum: significant absorption at $v \sim
60,000$~\kms\ is actually necessary to reproduce the observations.  We
then discuss the discrepancies with the later data, both light curve
and spectra, and suggest that asymmetry may be playing a key role.

\subsection{Model light curves}\label{sec:98bwlc}


\begin{figure}
\centerline{\psfig{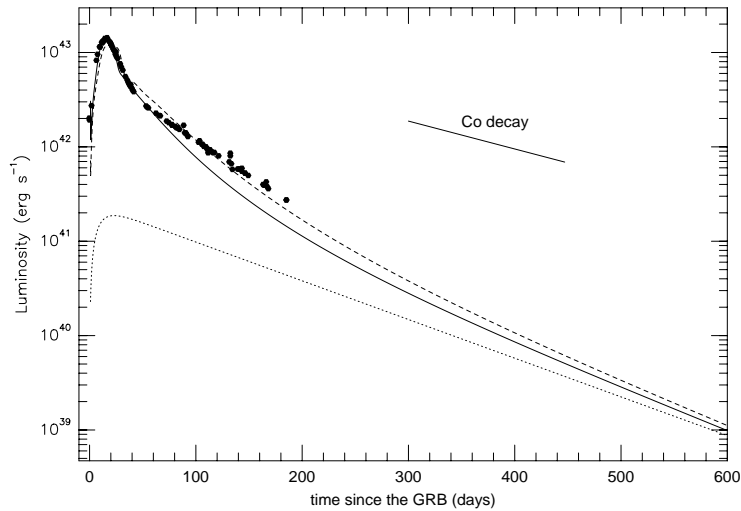}}
\caption{The light curves of models
CO138H ($E_{\rm K} = 6 \times 10^{52}$ erg; solid) and 
CO138L ($E_{\rm K} = 3 \times 10^{52}$ erg; dashed)
compared with the observations of SN1998bw
(Galama et al. 1999; McKenzie \& Schaefer 1999).
A distance modulus of $\mu = 32.89$ mag and $A_V = 0.0$ are adopted.
The dotted line indicates the energy deposited by positrons for CO138H.
\label{fig:98bwlc}}
\end{figure}

\begin{figure}
\centerline{\psfig{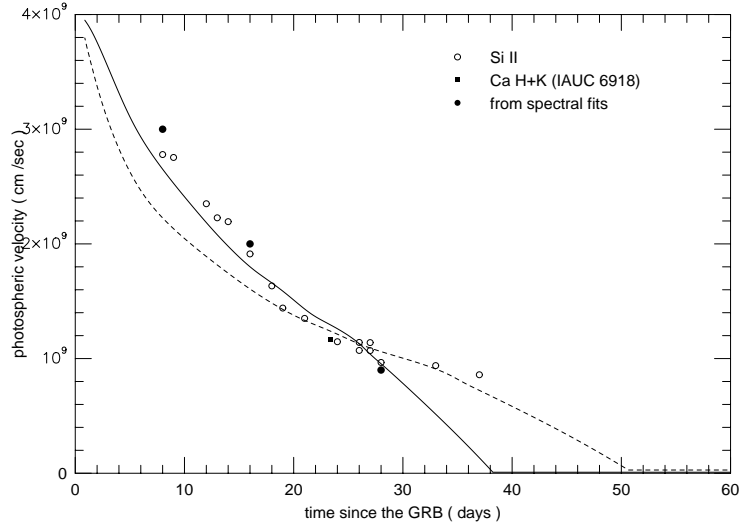}}
\caption{Photospheric velocities of models CO138H and CO138L
compared with the observations of SN1998bw.
\label{fig:98bwvph}}
\end{figure}

IMN98 modeled the SN~1998bw as the energetic explosion of a massive C+O
core, and obtained a good fit to the SN light curve in the first 60
days. The model was selected from a set of degenerate models because it
gave the best fit to the observed velocity of the \SiII\ line and
yielded sufficiently broad-lined spectra.

Nevertheless, the lines in the synthetic spectra were still noticeably
narrower than the observed features (IMN98, Fig.2). Also, photometry
after day $\sim 60$ show that SN~1998bw declined significantly more
slowly than the rate predicted by the model. Therefore, we recompute
the light curve of SN~1998bw using progenitors of different masses and
explosions of different energies. All models are spherically
symmetric. The ejected \Nifs\ is assumed to be rather centrally
distributed, and its mass was determined by fitting the light curve
around maximum.  We assume that the progenitor was a C+O star, and
searched a wide range of parameters.

We find that the model that give the best agreement to both the light
curve and the spectra is that of the explosion of a $13.8 \Msun$ C+O
star, ejecting 10 $\Msun$ of material with \KE~$ = 6 \times
10^{52}$~erg, including $0.5 \Msun$ of \Nifs~(CO138H).  In Figures
\ref{fig:98bwlc} and \ref{fig:98bwvph} we compare the bolometric light
curves and the photospheric velocities of model CO138H (solid) with
the V photometry of SN~1998bw. We use $\mu = 32.97$~mag, $A_V =
0.0$~mag, and assume that $BC = 0.0$.  This model has the same mass as
that published in IMN98, but it has a larger \KE, which is necessary
to improve the fit to the spectra (\S \ref{sec:98bwspe}).

However, CO138H has difficulties reproducing the apparently
exponential decline after day 60.  On the other hand, CO138L ($E_{\rm
K} = 3 \times 10^{52}$; dashed) is in better agreement after day 90,
although the early light curve and photospheric velocities do not fit
well.

After day $\sim$ 200 the decline of the model light curve becomes
slower, and it approaches the half-life of $^{56}$Co decay around day
400.  At $ t \gsim 400$ days most $\gamma$-rays escape from the
ejecta, while positrons emitted from the $^{56}$Co decay are mostly
trapped and their energies are thermalized.  Therefore, positron
deposition determines the light curve at $ t \gsim 400$ days (dotted
line in Fig. \ref{fig:98bwlc}).  If the observed tail should follow
the positron-powered light curve, the $^{56}$Co mass could be
determined directly.

The comparison between SN 1998bw and the model light curve of CO138H
(which fits better at early phases) and CO138L (which is better for
late phases) in Figure \ref{fig:98bwlc} suggests that there is a
significant amount of mass expanding at very low velocity, containing
some \Nifs. In this case the trapping time for the $\gamma$-rays would
be quite long, and the light curve might be explained. Indications for
a low-velocity, high-density region are found for SN~1997ef, a
lower-energy analogue of SN~1998bw.

This might indicate a departure from spherical symmetry. We will
discuss this further in the next section in combination with the
spectra. In any case, it is clear that the hydrodynamical model
obtained from the explosion has to be altered.

\subsection{Early Time Spectra}\label{sec:98bwspe}

\begin{figure}
\centerline{\psfig{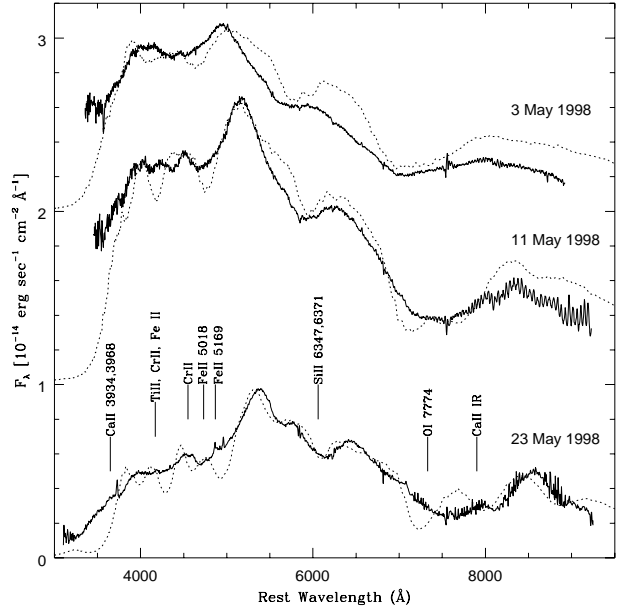}}
\caption{Observed spectra of SN1998bw (full lines) and synthetic spectra
calculated using model CO138H
($E_{\rm K} = 6 \times 10^{52}$~erg; dashed lines).
\label{fig:98bwspc}}
\end{figure}

We have used model CO138H as a basis to compute synthetic spectra for
the near-maximum phase of SN~1998bw. In particular, the model density
structure and composition were used as input into the Monte Carlo code
described above.

In Figure \ref{fig:98bwspc} we show the synthetic spectra obtained for
the same 3 epochs fitted in IMN98. The observed spectra used here are
the `definitive', fully reduced version of the same ESO spectra shown in
IMN98, and are calibrated with respect to the V photometry. The residual
correction factors for the other bands are usually very close to 1, but
they are $\sim 1.1$ for B in the May 11 and 23 spectra. Therefore, the
new models have somewhat different parameters than those of IMN98.  Both
the luminosity and the photospheric velocity have increased somewhat.
The photospheric velocity now is in even better agreement with the
measured velocity of the \SiII\ line (IMN98, Fig.3). All spectra are
computed assuming $\mu = 32.97$~mag and $A_V = 0.0$. The assumption of
zero reddening is supported by the upper limit of 0.1~\AA\ in the
equivalent width (EW) of the \NaI~D line obtained from high-resolution
spectra (Patat \etal 2000).

The synthetic spectra clearly improve over those of IMN98.  In
particular, those absorptions not due to broad blends, \ie the \SiII\
feature near 6000\AA, and the \OI+\CaII\ feature between 7000 and
8000\AA\ are now much broader, in significantly better agreement with
the data.  Nevertheless, the blue sides of those absorptions are still
too narrow, indicating that even the new model may not contain enough
mass at the highest velocities. Therefore we introduced an arbitrary
change to the original CO138 density structure. Several possibilities
were tested, and satisfactory results were found when the density slope
was reduced from $\rho \propto r^{-8}$ to $\rho \propto r^{-6}$ at $v >
30,000$~\kms. This does not introduce a significant change in \Mej, and
increases \KE\ by only about 10\%, but it does increase the density at
high velocities, leading to significant absorption at $v \sim
60,000$~\kms in the strongest lines, especially the \CaII~IR triplet,
extending the absorption troughs to the blue. The corresponding
synthetic spectra are shown as the dotted lines in Figure
\ref{fig:98bwspc}.  The effect of the change is largest at the earliest
epochs. The overall agreement with the observed spectra is better,
although several problems remain, the most severe of which is clearly
the excessive strength of the \OI\ line at 7200\AA\ on May 11 and 23.
The composition is dominated by O, and it is difficult to make that line
become weaker. On 23 May, the synthetic \CaII~IR triplet matches the
weak feature at 8000\AA, which is first seen on 11 May and which
continues to grow until it finally causes the wavelength of the
absorption minimum of the entire broad feature to shift to $\sim
8200$\AA\ (Patat \etal 2000). This is rather a peculiar
behavior, because on 3 May the \OI\ and \CaII\ lines had to blend much
more to give rise to the observed broad feature, which then had a
minimum at 7000\AA.

A very flat ($\rho \propto r^{-2}$) density distribution was also used
by Branch (2000) to fit the spectrum of SN~1998bw. This dependence is
however too flat when we use our MC model, because the ionization of \eg
\CaII\ does not fall as steeply as assumed by Branch (2000). On the other
hand, Branch's \KE\ ($5 \times 10^{52}$ erg) is similar to ours, but he
quotes a mass of $6 \Msun$ above 7000~\kms, while in our case the mass
above that velocity is as large as $\sim 10 \Msun$.

Clearly, a definitive solution has not been found yet. It is quite
possible that only by taking into account departures from spherical
symmetry it will be possible to get a really accurate fit to the
spectra. Nevertheless, considering the complexity of the problem, our
fits at least demonstrate that a large \KE\ is necessary, and that a
Type Ic SN O-dominated composition yields quite a reasonable
reproduction of the observations.

\subsection{Late Time Evolution}

\begin{figure}
\centerline{\psfig{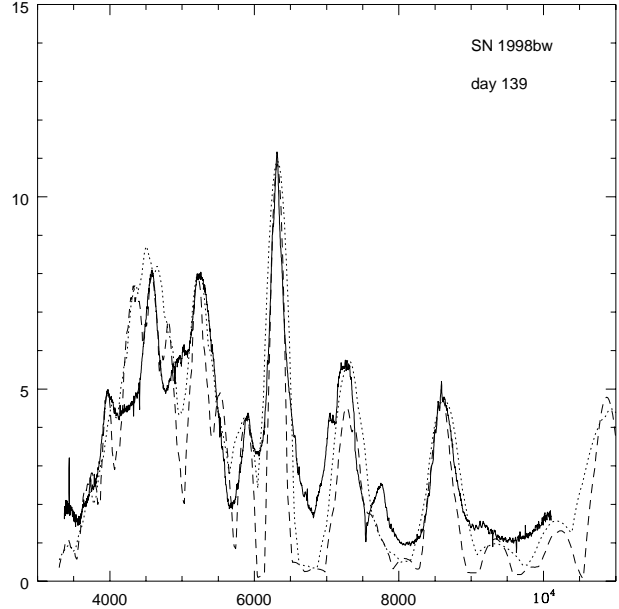}}
\caption{A nebular spectrum of SN~1998bw on 12 Sept 1998 (rest frame epoch 139
days) is compared to synthetic spectra obtained with a NLTE nebular model
based on the deposition of gamma-rays from $^{56}$Co decay in a nebula of
uniform density. Two models were computed.  In one model (dotted line) we tried
to reproduce the broad FeII] lines near 5300\AA.  The $^{56}$Ni mass is 0.65
$M_\odot$, and the outer nebular velocity is 11,000\kms, and the O mass is
3.5$M_\odot$. The average electron density in the nebula is log $n_e$ = 7.47
cm$^{-3}$.  In the other model (dashed line), we tried to reproduce only the
narrow [OI] 6300\AA\ emission line. These model has smaller $^{56}$Ni mass
(0.35 $M_\odot$) and O mass (2.1 $M_\odot$), and an outer velocity of
7500\kms.  The density is similar to that of the 'broad-lined' model.  The
filling factor used is 0.1 for both models.
\label{fig:98bwspclt}}
\end{figure}

Thus far we have presented the results of an analysis of SN~1998bw based on the
spherically symmetric model CO138H, which explain the observations around the
maximum. However, peculiarities in the spectrum began to appear soon after
maximum, when the broad feature in the red appeared to show a `double' \CaII\
absorption, or possibly an \OI\ absorption at much lower velocity than
predicted by the spherically symmetric models, which place O at the highest
velocity.

Following the evolution of the spectra as they become nebular, we see that the
SN showed a `composite' spectrum (Fig. \ref{fig:98bwspclt}; Danziger et al.
1999; Patat et al. 2000): \FeII] lines, typical of SNe~Ia, were strong, and so
were lines of \OI] and \MgI], which are typical of SNe~Ib/c. At the same time,
\FeIII] lines, also typical of SNe~Ia, were absent. The \OI] and \MgI] lines
grew stronger with time relative to the \FeII] lines, but had a narrower
profile, maybe the composite of a broad and a narrow profile, which dominates
more and more with time. The emergence of the narrow profiles occurs at about
the same time as the light curve deviates from the model prediction of CO138H.
In Fig. \ref{fig:98bwspclt} we show how one nebular spectrum can be
reproduced by two alternative models, one trying to fit the broad lines, and
the other aimed at fitting the narrower \OI] line. The \Nifs\ mass estimated
from the broad-line fit is comparable to the value obtained from the light
curve calculations. 

We suggest that a rather large mass of O-dominated material is also
present at low velocity. Spherical explosions do not allow that: they
are very effective at `emptying' the central region, and always place
the unburned elements at the top of the ejecta. So we suggest that the
explosion was highly asymmetric, leaving large quantities of unburned
material expanding at low velocity in directions away from the axis
along which most of the energy was released and \Nifs\
synthesized. Our vantage point must have been very close to that axis,
because we also detected the GRB.  At early times, the fast-expanding
lobes were much brighter than the rest, and so we observed the
broad-lined spectra and the bright light curve. The \Nifs\ mass
estimate of $\sim 0.6 M_\odot$ should not change much if the explosion
was not spherical. The fast-moving regions rapidly became thin, though,
and soon emission lines appeared. Initially those were broad, dominated
by the hyper-energetic lobes.  Later, though, the $\gamma$-rays from the
fast-moving \Cofs\ could escape that region more and more easily, and
a significant fraction of them could penetrate down into the
low-velocity region and excite the O and Mg there. Some \Nifs\ may
also be present in the low-velocity region, but that could only be
determined with better S/N observations of the nebular spectrum, which
require 8m-class telescopes at this point. Additional $\gamma$-ray
deposition in the low-velocity region may in turn increase the
deposition function above what our spherically symmetric model
estimates, and thus explain the slowly declining tail of the light curve.

Both the need for a high density region and the velocity inversion as
well as polarization measurements (Patat et al. 2000) might indicate
that the explosion is aspherical.  If the outburst in SN~1998bw took
the form of a prolate spheroid, for example, the explosive shock along
the long axis was probably strong, ejecting material with large
velocities and producing abundant $^{56}$Ni.  In directions away from
the long axis, on the other hand, oxygen is not much burned and the
density is high enough for $\gamma$-rays to be trapped even at advanced
phases, thus giving rise to the slowly declining tail.

That the SN~1998bw explosion was asymmetric is not a new suggestion:
first the polarization measurements indicated an axial ratio of 2-3:1
(H\"{o}flich et al. 1999) and the calculation of the explosion of a
rotating core (MacFadyen \& Woosley 1999) also gave similar results, in
an effort to explain the connection between SN~1998bw and
GRB980425. More detailed results have to await detailed numerical models
in two dimension. For the time being we can comment that if the
explosion was asymmetric, most likely our results for \KE\ are
overestimated, because those would only refer to the fast-moving part of
the ejecta. As for the value of \Mej, this can only be determined via 3D
hydrodynamical models of the explosion, but we hope that careful
analysis of the spectra, especially at late times, when both the fast
and the slow components are observable, can yield at least some
preliminary results.

We note that our estimate of the $^{56}$Ni mass of $\sim 0.6 M_\odot$ from the
nebula spectra in Figure \ref{fig:98bwspclt} does not much depend on the
asphericity. This is in good agreement with the spherical models CO138.  Since
H\"oflich et al. (1999) suggested that the $^{56}$Ni mass can be as small as
0.2 $M_\odot$ if aspherical effects are large, our results suggest that the
aspherical effects might be modest in SN~1998bw.

\section{Type IIn SN~1997cy}

\begin{figure}[t]
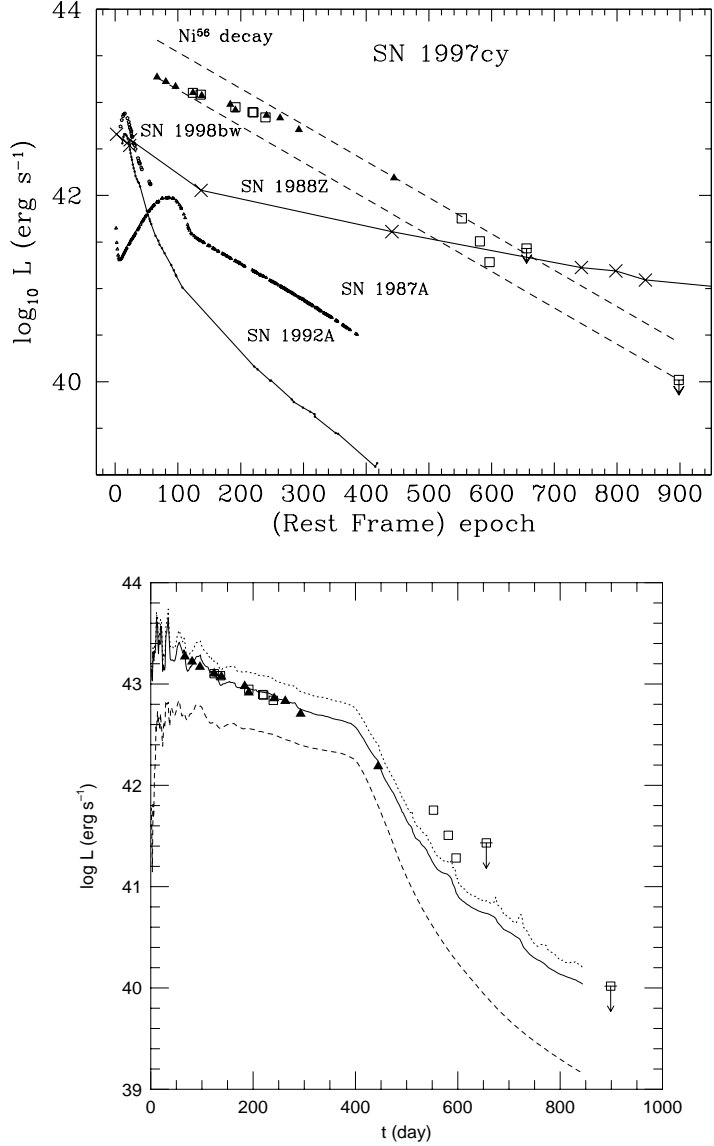

\centerline{
\hspace*{-2.cm}
\psfig{figure=97cynewlc.epsi,width=8cm}}
\vspace*{.5cm}
\centerline{\psfig{figure=97cylc2.epsi,width=8cm}}
\caption{
Upper panel: Observed light curves for SNe 1997cy, 1988Z,
1987A, 1992A, \& 1998bw.
Lower panel: The {\sl uvoir}  bolometric light curve of SN~1997cy
compared with the synthetic light curve obtained with the CSM
interaction model.
\label{fig:97cylc}}
\end{figure}

SN 1997cy displayed narrow H$\alpha$ emission on top of broad wings,
which lead to its classification as a Type IIn (Germany et al. 1999;
Turatto et al. 1999, 2000).  Assuming $A_V=0.00$ for the galactic extinction
(NED) we get an absolute magnitude at maximum $M_v
\le-20.1$.  It is the brightest SN II discovered so far.  The light
curve of SN 1997cy does not conform to the classical templates of SN
II, namely Plateau and Linear, but resembles the slow evolution of the
Type IIn SN 1988Z.  As seen from the {\it uvoir} bolometric light
curve in Figure \ref{fig:97cylc}, the SN light curve decline is slower
than the $^{56}$Co decay rate between day 120 to 250, suggesting
circumstellar interaction for the energy source.  (Here the outburst
is taken to be coincident with GRB970514.)

In the interaction model, collision of the SN ejecta with the slowly
moving circumstellar matter (CSM) converts the kinetic energy of the
ejecta into light, thus producing the observed intense light display
of the SN.  Our exploratory model considers the explosion of a massive
star of $M = 25 M_\odot$ with a parameterized kinetic energy $E_{\rm
K}$.  We assume that the collision starts near the stellar radius at a
distance $r_1$, where the density of the CSM is $\rho_1$, and adopt
for the CSM a power-law density profile $\rho \propto r^{n}$.  The
parameters $E_{\rm K}$, $\rho_1$, and $n$, are constrained from
comparison with the observations.

The regions excited by the forward and reverse shock emit mostly
X-rays. The density in the shocked ejecta is so high that the reverse
shock is radiative and a dense cooling shell is formed (e.g., Suzuki
\& Nomoto 1995; Terlevich et al. 1992).  The X-rays are absorbed by
the outer layers and the core of the ejecta, and re-emitted as
UV-optical photons.

Narrow lines are emitted from the slowly expanding unshocked CSM
photoionized by the SN UV outburst or by the radiation from the shocks;
intermediate width lines come from the shock-heated CSM; broad lines
come from either the cooler region at the interface between ejecta and
CSM.

Figure \ref{fig:97cylc} shows the model light curve which best fits
the observations.  The model parameters are: $E_{\rm K} =
3\times10^{52}$ erg, $\rho_1 = 4\times10^{-14}$ g cm$^{-3}$ at $r_1 =
2 \times10^{14}$ cm (which corresponds to a mass-loss rate of
$\dot{M}=4\times10^{-4}$ $M_{\odot}$ yr$^{-1}$ for a wind velocity of
10 \kms), and $n = -1.6$.  The large mass-loss episode giving rise to
the dense CSM is supposed to occur after the progenitor makes a loop
in the HR diagram from BSG to RSG.  In this model, the mass of the
low-velocity CSM is $\sim 5 M_\odot$, which implies that the
transition from BSG to RSG took place about $10^4$ yr before the SN
event.

The large CSM mass and density are necessary to have large shocked
masses and thus to reproduce the observed high luminosity, and so is
the very large explosion energy.  For models with low $E_{\rm K}$ and
high $\rho_1$, the reverse shock speed is too low to produce a
sufficiently high luminosity.  For example, a model with $E_{\rm K} =
10^{52}$ erg and $\rho_1$ as above yields a value of $L_{\rm UVOIR}$
lower than the observed luminosity by a factor of $\sim$ 5. For high
$E_{\rm K}$ or low $\rho_1$, the expansion of the SN ejecta is too
fast for the cooling shell to absorb enough X-rays to sustain the
luminosity.  Thus in this model $E_{\rm K}$ and $\dot{M}$ are
constrained within a factor of $\sim$ 3 of the reported values.

The shape of the light curve constrains the circumstellar density
structure. For $n = -2$, the case of a steady wind, $L_{\rm UVOIR}$
decreases too rapidly around day 200. To reproduce the observed
decrease after day $\sim$ 300, the CSM density is assumed to drop
sharply at the radius the forward shock reaches at day 300, so that
the collision becomes weaker afterwards.  (Such a change of the CSM
density corresponds to the transition from BSG to RSG of the
progenitor $\sim 10^4$ yr before the SN explosion.)  This is
consistent with the simultaneous decrease in the H$\alpha$
luminosity.


The observed light curve drops sharply after day 550.  We reproduce
such a light curve behavior (Figure \ref{fig:97cylc}) assuming that
when the reverse shock propagates through $\sim$ 5 $M_\odot$, it
encounters exceedingly low density region and thus it dies.  In other
words, the model for the progenitor of SN 1997cy assumes that most of
the core material has fallen into a massive black hole of, say, $\sim
10 M_\odot$, while the extended H/He envelope of $\sim$ 5 $M_\odot$
has not collapsed.  Then material is ejected from the massive black
hole possibly in a jet-like form, and the envelope is hit by the
``jet'' and ejected at high velocity.

In this model, the ejecta are basically the H/He layers and thus contain 
the original (solar abundance) heavy elements plus some heavy
elements mixed from the core (before fall back) or jet materials.  This
might explain the lack of oxygen and magnesium lines in the spectra
particularly at nebular phases (Turatto et al. 1999, 2000).

\section{Possible Evolutionary Scenarios to Hypernovae}

Here we classify possible evolutionary paths leading to C+O star
progenitors.  In particular, we explore the paths to the progenitors
that have rapidly rotating cores with a special emphasis, because the
explosion energy of hypernovae may be extracted from rapidly rotating
black holes (Blandford \& Znajek 1977).

(1) Case of a single star: If the star is as massive as $M_{\rm ms}
\gsim$ 40 \ms, it could lose its H and He envelopes in a strong stellar
wind (e.g., Schaller \etal 1992).  This would be a Wolf-Rayet star.

(2) Case of a close binary system: Suppose we have a close binary
system with a large mass ratio. In this case, the mass transfer from
star 1 to star 2 inevitably takes place in a non-conservative way, and
the system experiences a common envelope phase where star 2 is
spiraling into the envelope of star 1.  If the spiral-in releases
enough energy to remove the common envelope, we are left with a bare
He star (star 1) and a main-sequence star (star 2), with a reduced
separation.  If the orbital energy is too small to eject the common
envelope, the two stars merge to form a single star (e.g., van den
Heuvel 1994).

(2-1) For the non-merging case, possible channels from the He stars to
the C+O stars are as follows (Nomoto, Iwamoto, \& Suzuki 1995).

(a) Small-mass He stars tend to have large radii, so that they can
fill their Roche lobes more easily and lose most of their He envelope
via Roche lobe overflow.

(b) On the other hand, larger-mass He stars have radii too small to
fill their Roche lobes.  However, such stars have large enough
luminosities to drive strong winds to remove most of the He layer
(e.g., Woosley, Langer, \& Weaver 1995).  Such a mass-losing He star
would corresponds to a Wolf-Rayet star.

Thus, from the non-merging scenario, we expect two different kinds of
SNe Ic, fast and slow, depending on the mass of the progenitor.  SNe
Ic from smaller mass progenitors (channel 2-1-a) show faster light-curve
and spectral evolutions, because the ejecta become more quickly
transparent to both gamma-ray and optical photons. The slow SNe Ic
originate from the Wolf-Rayet progenitors (channels 1 and 2-1-b).  The
presence of both slow and fast SNe Ib/Ic has been noted by Clocchiatti
\& Wheeler (1997).

(2-2) For the merging case, the merged star has a large angular
momentum, so that its collapsing core must be rotating rapidly.  This
would lead to the formation of a rapidly rotating black hole from which
possibly a hyper-energetic jet could emerge.  If the merging process is
slow enough to eject the H/He envelope, the star would become a rapidly
rotating C+O star.  Such stars are the candidates for the progenitors of
Type Ic hypernovae like SNe 1997ef and 1998bw.  If a significant amount
of H-rich envelope remains after merging, the rapidly rotating core
would lead to a hypernova of Type IIn possibly like SN 1997cy (or Type Ib).

\section{Nucleosynthesis in Hypernovae}\label{sec:nuc}

Since hypernovae explode with much higher explosion energies than
usual supernovae, explosive nucleosynthesis could have some special
features.  Also hypernovae have shown some aspherical signatures.
Here we investigate how the explosive nucleosynthesis results depend
on the explosion energy and asphericity (Nomoto et al. 1998; Nakamura
et al. 1999c, 2000; Maeda et al.  2000).

\subsection{Nucleosynthesis in Spherical Explosions}

\begin{figure}
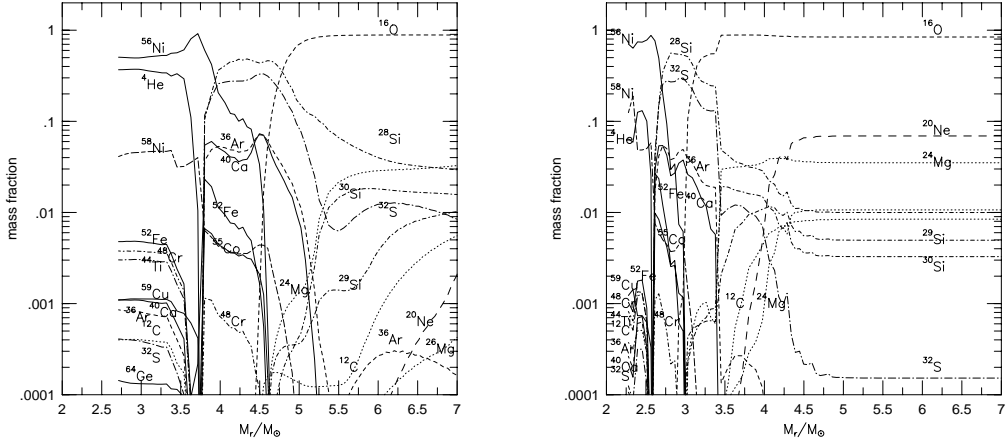

\hspace*{-.5cm}
\psfig{file=nssph1.epsi,width=6.cm}
\hspace*{1.cm}
\psfig{file=nssph2d.epsi,width=6.cm}
\caption{
The isotopic composition of ejecta of the hypernova
($E_{\rm K} = 3 \times 10^{52}$ erg; left)
and the normal supernova
($E_{\rm K} = 1 \times 10^{51}$ erg; right)
for a 16$M_\odot$ He star.
Only the dominant species are plotted.
The explosive nucleosynthesis is calculated
using a detailed nuclear reaction network including a total of 211
isotopes up to $^{71}$Ge.
\label{fig:nssph}}
\end{figure}

We calculate explosive nucleosynthesis in hypernovae in the same way
as has been done for normal supernovae; we use a detailed nuclear
reaction network including 211 isotopes up to $^{71}$Ge (Thielemann,
Nomoto, \& Hashimoto 1996; Hix \& Thielemann 1996; Nakamura et
al. 1999a) (Figure \ref{fig:nssph}: left).  Nucleosynthesis in normal
supernovae ($E_{\rm K} = 1 \times 10^{51}$ erg) is also shown in
Figure \ref{fig:nssph} (right) for comparison.

\begin{figure}
\hspace*{-1.cm}
\psfig{figure=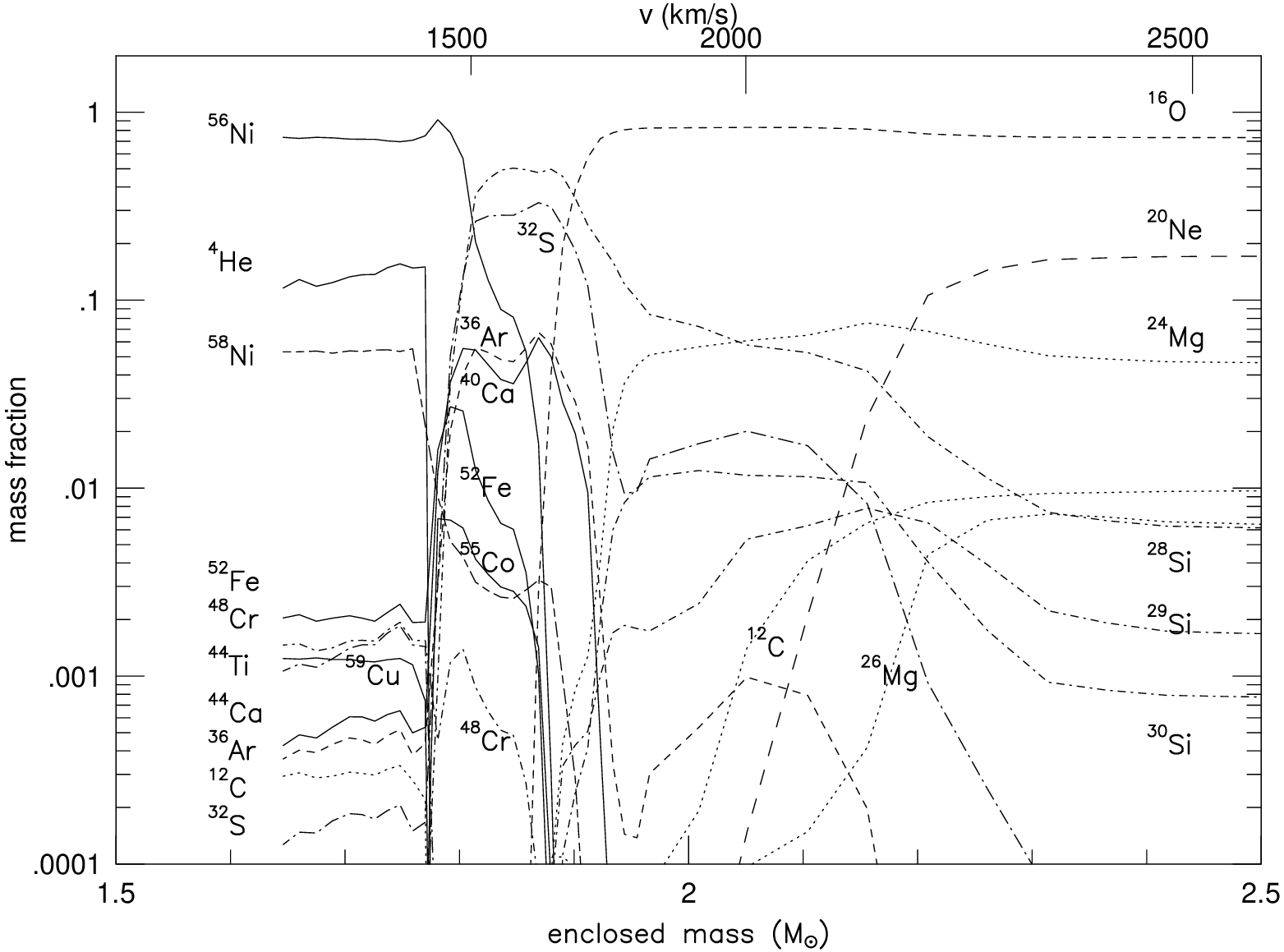,width=5.9cm}
\hspace*{.8cm}
\psfig{figure=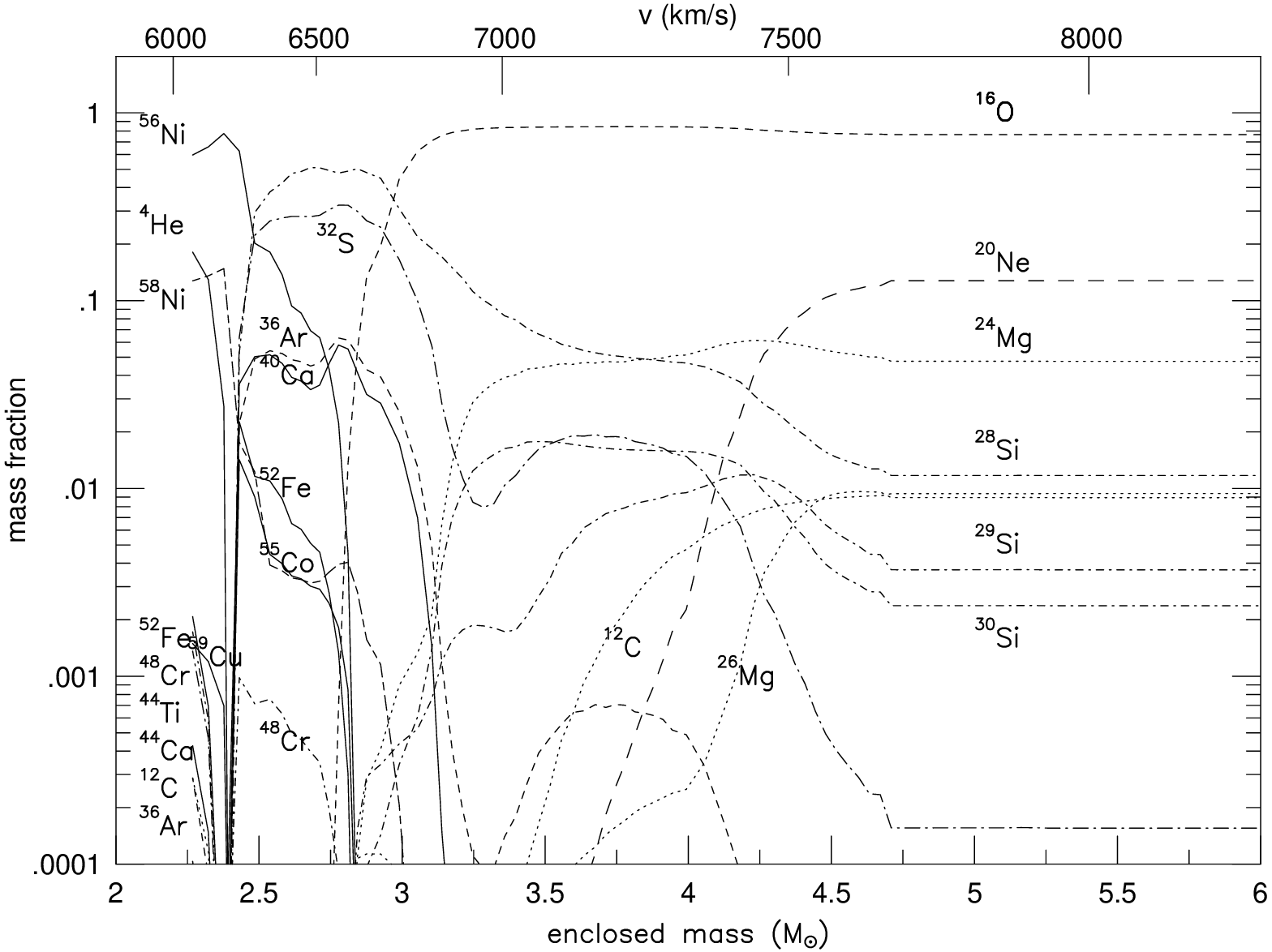,width=5.9cm}
\caption{The composition structure of 
models CO60 (left) and CO100 (right).
\label{fig:97efns}}
\end{figure}

A similar comparison is made in Figure \ref{fig:97efns}, which shows the
composition structure of models CO60 and CO100 against the expansion
velocity and the Lagrangian mass coordinate of the progenitor.  In
CO100, the Fe and Si-rich layers expand much faster than in CO60.  The
total amount of nucleosynthesis products are summarized in Table 2.

\begin{figure}
\centerline{\psfig{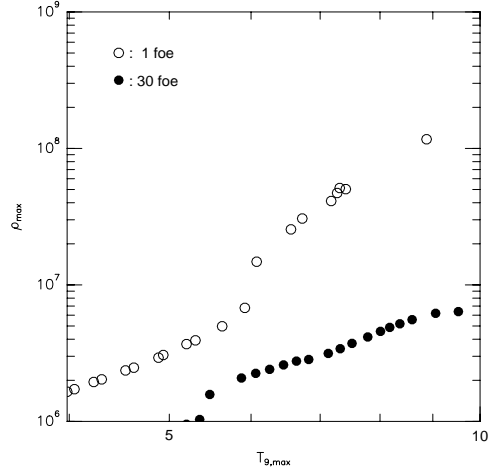}}
\caption{
The maximum $\rho$ - $T$ conditions of individual mass zones in the
normal supernova (\KE\ $=$ 1 foe $=$ $ 1 \times 10^{51}$ erg) and the
hypernova (\KE\ $=$ 30 foe).
\label{fig:maxrt}}
\end{figure}

\begin{figure}
\hspace*{-.5cm}
\psfig{file=nscompsun1.epsi,width=6.3cm}
\hspace*{.5cm}
\psfig{file=nscompsun2.epsi,width=6.3cm}
\caption{
Abundances of stable isotopes relative to the solar values
for 3 $\times 10^{52}$ ergs and 1 $\times 10^{51}$ ergs
(Figure \ref{fig:nssph}).
The progenitor is a 16$M_{\odot}$ He star
(H-rich envelope is not included).
\label{fig:nscompsun}}
\end{figure}

\begin{figure}
\hspace*{.1cm}
\psfig{figure=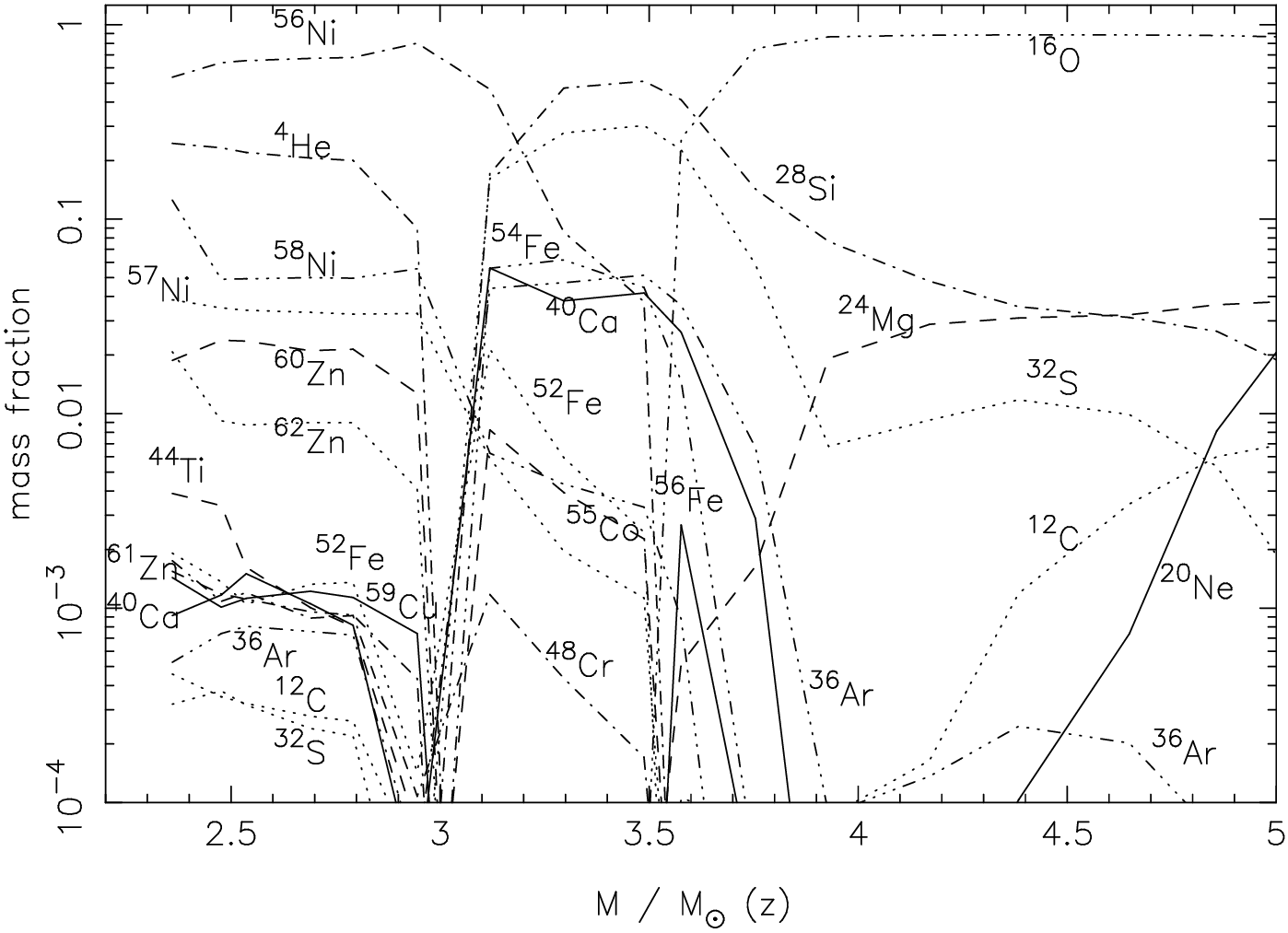,width=4.8cm}
\hspace*{2.cm}
\psfig{figure=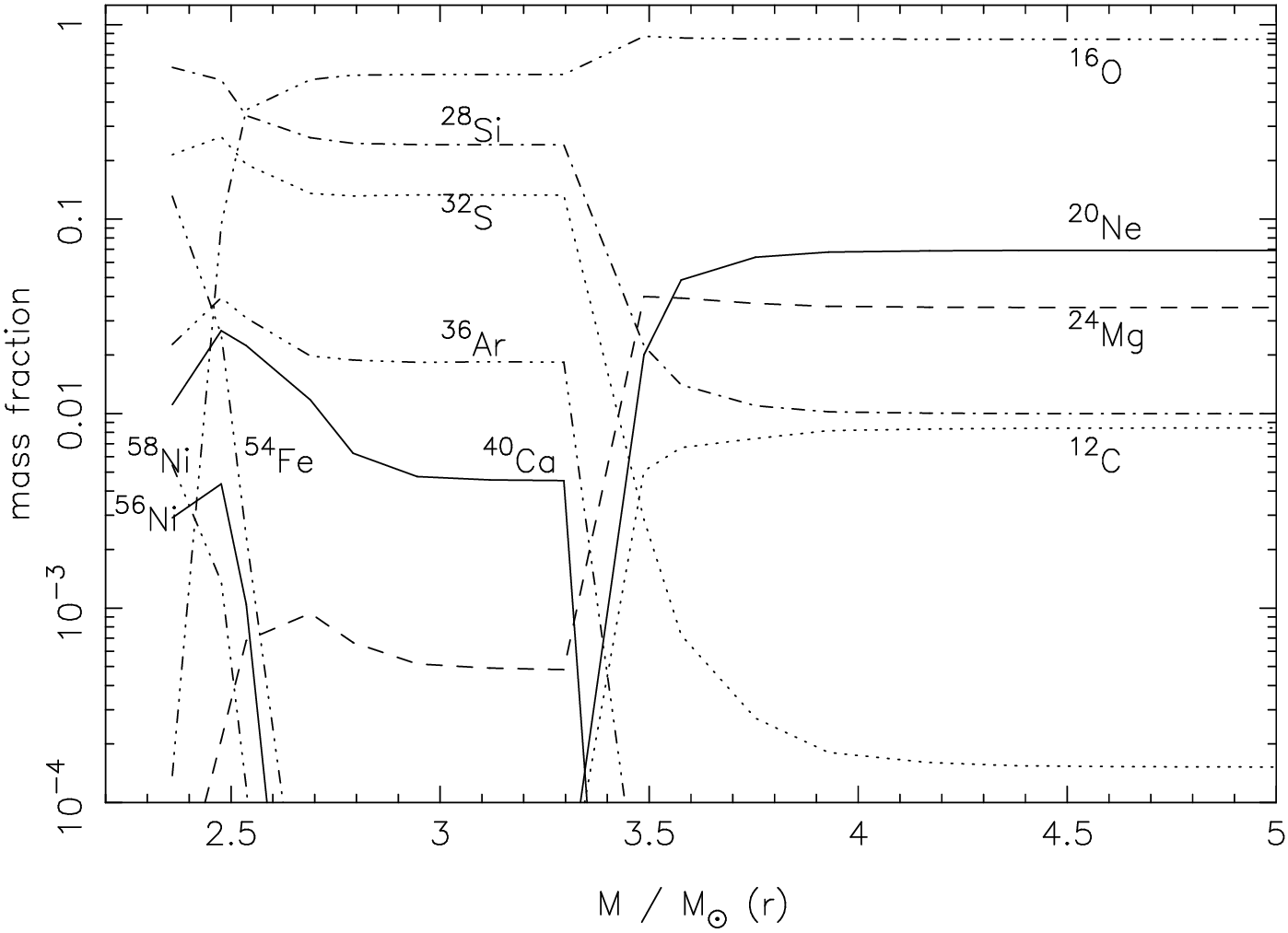,width=4.8cm}
\caption{
The isotopic compositions of the ejecta
of the axisymmetric explosion in the direction of
the jet (top) and the perpendicular to the jet (bottom)
with $E_{\rm K} = 1 \times 10^{51}$ erg (Maeda et al. 2000).
\label{fig:nsax}}
\end{figure}

\begin{table}
\begin{center}
\centerline{Table 2.~Yields of hypernova and supernova models ($M_\odot$)}
\vspace*{4mm}
\begin{tabular}{ccccccccccc} \hline
model& C & O & Ne & Mg & Si & S & Ca & Ti & Fe & Ni\\\
CO60& 0.082 & 3.0 & 0.62 & 0.24 & 0.10 & 0.037 & 0.006 & 0.0003& 0.16 &0.017\\\
CO100& 0.58 & 5.6 & 0.38 & 0.22 & 0.42 & 0.19 & 0.025 & 0.0003 & 0.19 & 0.021 \\
CO138H& 0.11 & 6.6 & 0.35 & 0.29 & 0.95 & 0.52 & 0.088 &0.0011 & 0.50 & 0.028\\
\hline\hline
model& $^{44}$Ti & $^{56}$Ni & $^{57}$Ni \\
CO60 & 2.1$\times 10^{-4}$ & 0.15 & 5.7$\times 10^{-3}$   \\
CO100& 4.5$\times 10^{-5}$ & 0.15 & 5.7$\times 10^{-3}$ \\
CO138H& 2.2$\times 10^{-4}$ & 0.50 & 1.5$\times 10^{-2}$ \\
\hline
\end{tabular}
\end{center}
\end{table}

 From this figure, we can see the following characteristics of
nucleosynthesis with the very large explosion energy.

1) The complete Si-burning region is extended to the outer, lower
density region.  Whether this region is ejected or not depends on the
mass cut.  The large amount of $^{56}$Ni observed in hypernovae (e.g.,
$\sim$ 0.5$M_{\odot}$ for SN1998bw and 0.15$M_{\odot}$ for SN1997ef)
implies that the mass cut is rather deep, so that the elements
synthesized in this region such as $^{59}$Cu, $^{63}$Zn, and $^{64}$Ge
(which decay into $^{59}$Co, $^{63}$Cu, and $^{64}$Zn, respectively)
are likely to be ejected more abundantly.  In the complete Si-burning
region of the hypernova, elements produced by $\alpha$-rich freezeout
are enhanced because nucleosynthesis proceeds under lower densities
than in usual supernovae (Fig. \ref{fig:maxrt}).  Figure
\ref{fig:nssph} clearly shows a trend that a larger amount of $^{4}$He
is left in more energetic explosion.  Hence, elements synthesized
through $\alpha$-captures such as $^{40}$Ca (stable), $^{44}$Ti and
$^{48}$Cr (decaying into $^{44}$Ca and $^{48}$Ti, respectively) become
more abundant.

2) The more energetic explosion produces a broader incomplete Si-burning
region.  The elements produced mainly in this region such as $^{52}$Fe,
$^{55}$Co, and $^{51}$Mn (decaying into $^{52}$Cr, $^{55}$Mn, and
$^{51}$V, respectively) are synthesized more abundantly with the larger
explosion energy.

3) Oxygen burning takes place in more extended, lower density region for
the larger explosion energy, so that the abundances of elements like O,
C, Al are smaller.  On the other hand, a larger amount of ash products
such as Si, S, Ar are synthesized by oxygen burning.

Figure \ref{fig:nscompsun} shows the abundances of stable isotopes
relative to the solar values for 3 $\times$ 10$^{52}$ erg and 1
$\times$ 10$^{51}$ erg.  The progenitor is the 16$M_{\odot}$ He star
and products from H-rich envelope are not included.  The isotopic
ratios relative to $^{16}$O with respect to the solar values are
shown.  As a whole, intermediate mass nuclei and heavy nuclei are more
abundant for the more energetic explosion, except for the elements
being consumed in oxygen burning like O, C, Al.  Especially, the
amounts of $^{44}$Ca and $^{48}$Ti are increased significantly because
of the enhanced $\alpha$-rich freezeout.

\subsection {Nucleosynthesis in Asymmetric Explosions}

In \S \ref{sec:98bw}, we speculate that the expansion velocity of Fe and
O in SN~1998bw betrays the effect of the asymmetry in the explosion.  To
confirm this we calculate the explosive nucleosynthesis in an
axisymmetric explosion.  Figure \ref{fig:nsax} shows the isotopic
compositions of the ejecta of the axisymmetric explosion in the
direction of the jet (left) and perpendicular to the jet (right).
The progenitor model is CO138.  The the explosion energy is $E_{\rm K} =
1 \times 10^{51}$ erg.  Starting the hydrodynamical simulation, we
deposit the energy as 50\% thermal energy and 50\% kinetic energy toward
the jet (z) below the mass cut that divides the ejecta and the
collapsing core.

The shock is stronger and the post-shock temperatures are higher along
the jet direction (z), so that explosive nucleosynthesis takes place in
a more extended, lower density region compared with the perpendicular
direction (r).  A larger amount of $^{56}$Ni is produced in the jet
direction.  In addition, elements produced by $\alpha$-rich freezeout
are enhanced because nucleosynthesis proceeds at higher entropies than
in the region away from the jet.  Figure \ref{fig:nssph} clearly shows
that in the jet direction a larger amount of $^{4}$He is left after the
shock decomposition.  Hence, elements synthesized through capturing
$\alpha$-particles such as $^{44}$Ti and $^{48}$Cr (decaying into
$^{44}$Ca and $^{48}$Ti, respectively) become more abundant (see also
Nagataki et al. 1997).  In contrast, little $^{56}$Ni is produced in the
r-direction.  Also the expansion velocities are lower than those in the
z-direction.  Therefore, the Fe velocities (mostly z-direction) can
exceed the O velocities (in the r-direction), as observed in SN~1998bw.
Oxygen in the z-direction has the highest velocities but the densities
may become too low to be excited by gamma-rays.

Such an asymmetric ejection of nucleosynthesis products may explain
the abundance features observed in X-ray Nova Sco (GRO J1655-40),
which consists of a massive black hole and a low mass companion (e.g.,
Brandt et al. 1995; Nelemans et al. 2000).  The companion star is
enriched with Ti, S, Si, Mg, and O but not much Fe (Israelian et
al. 1999). This is compatible with heavy element ejection from a black
hole progenitor.  In order to eject large amount of Ti, S, and Si and
to have at least $\sim$ 4 \ms\ below mass cut and thus form a massive
black hole, the explosion should be highly energetic
(Fig. \ref{fig:nssph}; Israelian et al. 1999; Brown et al. 2000a;
Podsiadlowski et al. 2000).  Suppose that an asymmetric explosion
occurred when the black hole formed in Nova Sco.  Then it is likely
that the companion star captured material ejected in the r-direction
(i.e., on the orbital plane) which contains relatively little Fe
compared with the z-direction, where burning is more effective
(Podsiadlowski et al. 2000; Brown et al. 2000b).  Quantitatively,
nucleosynthesis in the r-direction for \KE\ $=$ 1 \e{52} erg is in
good agreement with Nova Sco (Maeda et al. 2000).

\subsection{The Mass of Ejected \(^{56}\)Ni }

For the study of the chemical evolution of galaxies, it is important
to know the mass of $^{56}$Ni, $M(^{56}$Ni), synthesized in
core-collapse supernovae as a function of the main-sequence mass
$M_{\rm ms}$ of the progenitor star (e.g., Nakamura et al. 1999a).
 From our analysis of SNe 1998bw and 1997ef, we can add new points on
this diagram.

We evaluate the uncertainty in our estimates of $M(^{56}$Ni) and
$M_{\rm ms}$. We need 0.15 \ms\ of $^{56}$Ni to get a reasonable fit
to the light curve of SN 1997ef at a distance $D = 52.3$ Mpc.  The
expected 10\% uncertainty in the distance leads to a 20\% uncertainty
in the $^{56}$Ni mass, i.e., $M(^{56}$Ni) $=$ 0.15 $\pm$ 0.03 \ms.
The distribution of $^{56}$Ni affects the peak luminosity somewhat,
but the effect is found to be much smaller than that of the
uncertainty in the distance.  A 10 \ms\ C+O star corresponds to a
$M_{\rm ms} =$ 30 - 35 \ms, but the uncertainty involved in the
conversion of the core mass to $M_{\rm ms}$ may involve a larger
uncertainty if the progenitor undergoes close binary evolution.

\begin{figure}
\centerline{\psfig{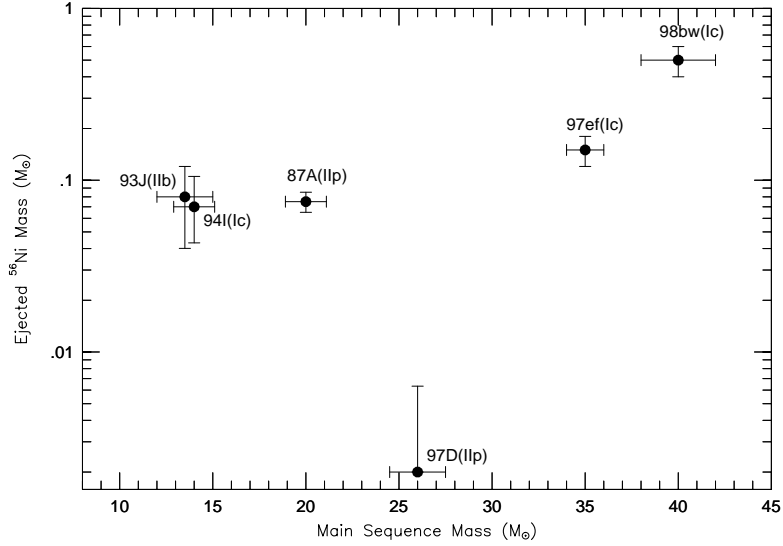}}
\caption{Ejected $^{56}$Ni mass versus the main sequence mass of the 
progenitors of several bright supernovae obtained from light curve models.
\label{fig:nimass}}
\end{figure}

Figure \ref{fig:nimass} shows $M(^{56}$Ni) against $M_{\rm ms}$ obtained
from fitting the optical light curves of SNe 1987A, 1993J, and 1994I
(e.g., Shigeyama \& Nomoto 1990; Nomoto et al. 1993, 1994; Shigeyama
\etal 1994; Iwamoto \etal 1994; Woosley et al. 1994; Young, Baron, \&
Branch 1995).  The amount of $^{56}$Ni appears to increase with
increasing $M_{\rm ms}$ of the progenitor, except for SN II 1997D
(Turatto et al. 1998).

This trend might be explained as follows.  Stars with $M_{\rm ms}
\lsim$ 25 \ms\ form a neutron star, producing $\sim$ 0.08 $\pm$ 0.03
\ms\ \ni\ as in SN IIb 1993J, SN Ic 1994I, and SN 1987A (although SN
1987A may be a borderline case between neutron star and black hole
formation). Stars with $M_{\rm ms} \gsim$ 25 \ms\ form a black hole
(e.g., Ergma \& van den Heuvel 1998); whether they become hypernovae
or ordinary SNe may depend on the angular momentum in the collapsing
core.  For SN 1997D, because of the large gravitational potential, the
explosion energy was so small that most of \ni\ fell back onto a
compact star remnant; the fall-back might cause the collapse of the
neutron star into a black hole.  The core of SN II 1997D might not
have a large angular momentum, because the progenitor had a massive
H-rich envelope so that the angular momentum of the core might have
been transported to the envelope possibly via a magnetic-field effect.
Hypernovae such as SNe 1998bw, 1997ef, and 1997cy might have rapidly
rotating cores owing possibly to the spiraling-in of a companion star
in a binary system.  The outcome certainly depends also on mass-loss
rate and binarity.

\section{Gamma-Ray Bursts/Supernovae Connection}

Candidates for the GRB/SN connection include GRB980425/SN Ic 1998bw
(Galama et al. 1998; IMN98), GRB971115/SN Ic 1997ef (Wang \& Wheeler
1998), GRB970514/SN IIn 1997cy (Germany et al. 1999; Turatto et
al. 1999, 2000), GRB980910/SN IIn 1999E (Thorsett \& Hogg 1999), and
GRB991002/SN IIn 1999eb (Terlevich et al. 1999).

Two other GRB's may also be associated with a SN: GRB980326 (Bloom et
al. 1999) and GRB970228 (Reichart 1999; Galama et al. 1999).  The
optical afterglows of these GRBs showed that the decline of the light
curve is slowed down at late phases, and this can be reproduced if a
red-shifted SN~1998bw-like light curve is superposed on the power-law
light component.  A question arising from these two examples is
whether the supernovae associated with GRBs have a uniform maximum
luminosity, i.e., whether $\sim$ 0.5 \ms\ \ni\ production as in SN
1998bw is rather common or not.  However, the present study of SN
1997ef shows that the \ni\ mass and thus intrinsic maximum brightness
of SN 1997ef is smaller than in SN 1998bw by a factor of 4 - 5.  We
certainly need more examples for defining the luminosity function and
the actual distribution of masses of \(^{56}\)Ni produced in
supernovae/hypernovae.

Among the possible connections suggested above, the statistical
significance for the case of SN~1997ef and GRB971115 is much weaker
than for the case of SN~1998bw and GRB980425.  Recently another SN Ic,
1998ey, showed a spectrum with very broad features, very similar to
that of SN1997ef on Dec 17 (Garnavich et al. 1998); but no GRB
counterpart has been proposed for SN~1998ey. Although this may cast
some doubt on the general association between hypernovae and GRBs, it
must be noted that both SNe 1997ef and 1998ey were less energetic
events than SN~1998bw. It is possible that a weaker explosion is less
efficient in collimating the $\gamma$-rays to give rise to a
detectable GRB (GRB980425 was already quite weak compared to the
average GRBs), or that some degree of inclination of the beam axis to
the line-of-sight results in a seemingly weaker supernova and in the
non-detection of a GRB. Only the accumulation of more data will allow
us to address these questions.

\section{Concluding remarks}

We have calculated the light curves and spectra for various C+O star
models with different values of $E_{\rm K}$ and $M_{\rm ej}$ and
reached several striking conclusions.  

We have shown that the spectra of SNe 1998bw and 1997ef are much
better reproduced with the hypernova models than with the ordinary SN
Ic model. Since SN1998bw was connected to a highly non-spherical event
like a GRB, departure from spherical symmetry could be expected.
Early polarization measurements confirmed this: polarization of $\sim
1$\%, decreasing with time, was detected.

The evidence for asphericity in SN 1998bw becomes even stronger with
the extended time coverage: the light curve decline is slower than
predicted by our spherically symmetric model; the composite nebular
spectra have different velocities in lines of different elements, with
iron expanding more rapidly than oxygen; the \OI] nebular line
declines more slowly than the \FeII\ ones, signaling deposition of
$\gamma$-rays in a slowly-moving O-dominated region.  

The smaller line velocities at advanced phases and the flat light
curve tail of SN~1997ef may also suggest the presence of a
low-velocity, relatively dense core, while the high line velocities at
early phases imply the presence of an even higher-velocity component
of the ejecta.  This discrepancy between models and observations, as
well as the extensive mixing of \(^{56}\)Ni required to explain the
early rise of the light curve, seems to indicate that the explosion of
SN 1997ef was at least somewhat aspherical.

Therefore, we suggest that SNe 1997ef, 1998ey, and 1998bw form a new
class of hyper-energetic Type Ic supernovae, which we may call ``Type
Ic'' hypernovae.  SN~1998bw produced $\sim$ 0.5 - 0.7$M_\odot$ of
$^{56}$Ni, as much as a SN Ia, while SN~1997ef produced less, only
$\sim$ 0.15$M_\odot$, but still more than in ordinary SNe~Ic.
SN~1997ef also appeared to be less energetic than SN~1998bw. This may
be a real difference, but it may also result from different
inclination or beaming properties, since no GRB counterpart was
positively observed for SN~1997ef.

SNe 1997cy, 1999E, and 1999eb may form a class of ``Type IIn''
hypernovae. They are also distinguished by their large kinetic
energies, 8 - 60 times larger than in ordinary supernovae, but it is
not easy to determine how much \Nifs\ they produced since their light
curves and spectra are dominated by interaction with a massive CSM.
Simulations of the interaction can reproduce the observed light curve
of SN~1997cy, indicating that the progenitor must have been a massive
star, which possibly underwent spiral-in of the companion star in a
close binary system.

Continuing observations and theoretical modeling of this interesting
class of objects are certainly necessary.

\medskip

This work has been supported in part by the grant-in-Aid for COE
Scientific Research (07CE2002) of the Ministry of Education, Science,
Culture and Sports in Japan.

\end{document}